\documentclass[prb,twocolumn,floatfix]{revtex4}

\usepackage{amsfonts}
\usepackage{amsmath}
\usepackage{verbatim}
\usepackage[dvips]{graphicx}
\usepackage[dvips]{color}
\usepackage{subfigure}

\def\dd{\textrm{d}}

\begin{document}

\title{Vortices and vortex states in Rashba spin-orbit-coupled condensates}
\author{Predrag Nikoli\'c$^{1,2}$}
\affiliation{$^1$School of Physics, Astronomy and Computational Sciences,\\George Mason University, Fairfax, VA 22030, USA}
\affiliation{$^2$Institute for Quantum Matter at Johns Hopkins University, Baltimore, MD 21218, USA}
\date{\today}


\begin{abstract}

The Rashba spin-orbit coupling is equivalent to the finite Yang-Mills flux of a static SU(2) gauge field. It gives rise to the protected edge states in two-dimensional topological band-insulators, much like magnetic field yields the integer quantum Hall effect. An outstanding question is which collective topological behaviors of interacting particles are made possible by the Rashba spin-orbit coupling. Here we addresses one aspect of this question by exploring the Rashba SU(2) analogues of vortices in superconductors. Using the Landau-Ginzburg approach and conservation laws, we classify the prominent two-dimensional condensates of two- and three-component spin-orbit-coupled bosons, and characterize their vortex excitations. There are two prominent types of condensates that take advantage of the Rashba spin-orbit coupling. Their vortices exist in multiple flavors whose number is determined by the spin representation, and interact among themselves through logarithmic or linear potentials as a function of distance. The vortices that interact linearly exhibit confinement and asymptotic freedom similar to quarks in quantum chromodynamics. One of the two condensate types supports small metastable neutral quadruplets of vortices, and their tiles as metastable vortex lattices. Quantum melting of such vortex lattices could give rise to non-Abelian fractional topological insulators, SU(2) analogues of fractional quantum Hall states. The physical systems in which these states could exist are trapped two- and three-component bosonic ultra-cold atoms subjected to artificial gauge fields, as well as solid-state quantum wells made either from Kondo insulators such as SmB$_6$ or conventional topological insulators interfaced with conventional superconductors.

\end{abstract}

\maketitle

\section{Introduction}

The discovery of topological insulators (TI) \cite{Kane2005a, Kane2005, Bernevig2006, Bernevig2006a, Fu2007, Konig2007, Moore2007, Teo2008, Fu2008, Chen2009b} has rejuvenated the longtime interest in the Rashba spin-orbit coupling. In addition to interfaces in solid state heterostructures, the Rashba spin-orbit coupling is naturally found on the boundaries of strong topological insulators where it creates a Dirac-like dispersion for protected electronic surface states \cite{Fu2007}. Extremely thin topological insulator films have a similar dynamics, but their Dirac electrons are gapped by the coupling between two nearby surfaces \cite{Murakami2004, Bernevig2006, Konig2007}. All firmly established TIs so far are uncorrelated band-insulators \cite{Konig2007, Hasan2010, Qi2010a, Moore2010}, but Kondo insulators such as samarium hexaboride (SmB$_6$) are emerging as likely topologically non-trivial materials with natural strong correlations among electrons \cite{Dzero2010, Dzero2012, Zhang2013, Wolgast2013, Kim2013, Kim2013a, Thomas2013}.

A significant effort to understand and experimentally create a Rashba spin-orbit coupling has recently taken place in the context of ultra-cold atoms \cite{Dudarev2004, Osterloh2005, Ruseckas2005, Zhu2006, Jacob2007, Liu2009b, Stanescu2009, Sau2011, Br2012, Galitski2013}. A major motivation is to mimic electronic systems using fermionic atoms \cite{Wang2012b, Cheuk2012}. However, cold atoms subjected to artificial gauge fields \cite{Spielman2009, Gunter2009, Juzeliunas2010, Dalibard2010, Campbell2011, Goldman2010, Br2013, Lan2014} can be used to engineer bosonic systems with the Rashba spin-orbit coupling, and explore their unconventional condensates unavailable in solid state systems. At this time, creating a proper Rashba spin-orbit coupling is an experimental challenge \cite{Campbell2011, Br2013}. The existing cold-atom implementations of spin-orbit coupling are either the equal combination of Rashba and Dresselhaus couplings which is topologically trivial \cite{Spielman2009, Lin2011, Wang2012b, Cheuk2012}, or a spin-dependent effective magnetic field which aims to produce quantum spin-Hall states \cite{Miyake2013, Kennedy2013, Beeler2013}. The Rashba spin-orbit coupling is capable of yielding more exotic physics. Many theoretical studies have sought such physics by studying the ground state of trapped spin-orbit-coupled bosons \cite{Stanescu2008, Wang2010b, Ho2011, Yip2011, Wu2011, Su2012, Kawakami2012, Xu2012b, Song2012, Zhai2012, Sinha2011a, Radic2011, Ruokokoski2012, Sedrakyan2012, Jian2011, Gopalakrishnan2011, Hu2012, Ozawa2011, Zhu2012, Barnett2012, Ozawa2012, Stringari2012, Zhou2013, Zhou2013a, Ozawa2013, Ramach2013, Riedl2013, Lu2013, He2014, Wilson2013, Demler2013}.

In this paper we pursue the quest for unconventional states of matter further, by exploring the nature of vortex states in the two-dimensional condensates of Rashba spin-orbit coupled particles. The same similarity that exists between two-dimensional TIs and quantum Hall systems applies to the comparison between the condensates we explore and superconductors in magnetic fields. Both the Rashba spin-orbit coupling and magnetic fields are mathematically represented by external static gauge fields with non-zero fluxes presented to particles \cite{Frohlich1992}. The former belongs to the SU(2) symmetry group, while the latter is U(1). This gives rise to various similar phenomena, but also to differences because the SU(2) group is non-Abelian. Plain condensates of Rashba coupled particles tend to be very good mean-field candidates for the ground state in the continuum. However, tight-binding lattice bosons with a strong Rashba spin-orbit coupling are capable of forming condensates with an entire vortex lattice \cite{Cole2012, Radic2012}, distantly reminiscent of the superconductor's Abrikosov lattice.

We are interested here in the fundamental theoretical question: what is the character of vortices and possible vortex lattices in the Rashba spin-orbit coupled condensates? We approach this problem within the Landau-Ginzburg description of two-dimensional interacting bosons at finite temperatures. In particular, we consider two- and three-component bosonic systems. The former are motivated in the context of cold-atoms \cite{Spielman2009, Campbell2011, Goldman2010, Br2013}, while the latter can arise effectively from electron pairing in correlated solid state quantum wells \cite{Nikolic2011a, Nikolic2012b}. In the first step we obtain the naive mean-field phase diagrams of these systems, and discover two prominent phases that gain Rashba energy via ``spin currents''. The phase we denote ``type-I'' establishes a pure spin current, while the phase we call ``type-II'' creates a spin current from the movement of particles with a definite spin orientation. The former has been also dubbed ``striped'' and the latter ``plain-wave'' state in the context of two-component bosons \cite{Ho2011, Wang2010b, Zhai2012, Ozawa2012}, but this terminology may not be equally descriptive in higher spin representations. Then, we study vortex excitations with loop currents in these condensates. Being multi-component, these systems harbor multiple flavors of vortices. There are two characteristic singular patterns of spin currents which we dub ``helical'' and ``chiral'', and a standard charge vortex that exists even in conventional superfluids. Three-component systems have two additional types of vortex singularities. Charge vortices are always found to interact among themselves through a logarithmic two-dimensional Coulomb potential. However, chiral vortices are very costly and interact among themselves through a linear potential as a function of distance. This gives them a dynamics similar to that of quarks in quantum chromodynamics. We identify the quantized vortex charges of all singularities and the qualitative current patterns in their vicinity.

The mean-field condensates are naively uniform, but in some cases have periodically modulated densities or currents \cite{Ozawa2012a} that break the translation symmetry. These modulations appear just at short length-scales and have a trivial topology of spin currents, so we only briefly discuss them in the appendix and otherwise ignore them as a nuisance of no consequence for the character of vortices. In any case, the mean-field condensates are qualitatively different from Abrikosov lattices in superconductors. An important discovery of our analysis is that the type-I condensates allow small classically metastable neutral clusters of vortices. The elementary cluster is a quadruplet of singularities that are kept at optimum distances from one another. Quadruplets can be tiled into an entire metastable lattice of vortices and antivortices \cite{Nikolic2011a}. Given this general metastability, microscopic systems in which such lattices are stable surely exist. Good candidates are bosons moving through tight-binding crystals, where small vortex cores can sit inside lattice plaquettes and have a low cost. The entropy of quantum fluctuations beyond the mean-field approximation further enhances the stability of vortex states in an order-by-disorder fashion. Vortex lattices of the kind we theoretically elucidate in this paper have already been seen in numerical studies of the Rashba coupled lattice bosons \cite{Cole2012, Nikolic2014a}.

A truly stable vortex lattice created by the Rashba SU(2) flux is more than just an unconventional symmetry broken state. It is also significant as a starting point toward realizing very exotic incompressible liquid states of quantum matter. The amount of the SU(2) Yang-Mills flux regulates the density of vortices in the predicted lattice state, much like the magnetic field sets the density of one vortex per flux quantum in the superconductors' Abrikosov lattices. Positional quantum fluctuations of vortices can melt the lattice if the superfluid stiffness is not large enough to support strong inter-vortex interactions. This happens when the number of condensed particles per vortex becomes small. Quantum melting of Abrikosov lattices has been argued to yield fractional quantum Hall liquids \cite{Wilkin2000, Cooper2001, Regnault2003, Chang2005, Cooper2008, nikolic:144507, Nikolic2009}. It is entirely possible by analogy that quantum melting of the type-I vortex lattice would also produce a fractional quantum liquid. Field-theoretical arguments \cite{Nikolic2012} strongly support this view, and predict that the ensuing quantum liquid would naturally have a novel kind of fractional quasiparticles with non-Abelian statistics, possibly amenable to quantum computing. Other proposals of exotic states of Rashba spin-orbit coupled bosons \cite{Sedrakyan2012} are perhaps related to this picture.

Given our focus on the fundamental phenomena in this paper, we do not discuss any practical means to realize the states we mention, or attempt to derive their detailed experimental manifestations. The phases we describe are bound to exist at low temperatures in any system whose low energy degrees of freedom are weakly interacting bosons. Such systems with artificial spin have already been engineered using cold atoms \cite{Spielman2009, Campbell2011}, although creating the pure Rashba spin-orbit coupling remains a challenge. Similarly, many solid-state materials exhibit electron pairing in either Cooper or exciton channels, which generates low-energy bosonic degrees of freedom. Pairing is particularly prominent in two spatial dimensions, where it can occur even in weakly coupled band-insulators \cite{Nikolic2010, Nikolic2010b, Nikolic2012b}. Triplet pairs in TI quantum wells would be energetically enhanced at large momenta by the Rashba spin-orbit coupling, and might become coherent bosonic excitations that can condense. Irrespective of these details, our goal is merely to predict the universal measurable consequences of condensation in the presence of the Rashba spin-orbit coupling. For example, understanding vortex excitations and their interactions in the ``uniform'' superfluids is important for predicting their unbinding phase transitions at finite temperatures. Some of the vortices we find are confined by a linear potential; their unbinding transitions can be detected by thermodynamic probes, but need not be in the well-known Kosterlitz-Thouless universality class. This will be the subject of a future study. Similarly, the vortex lattice state we predict will be easily observable if it emerges in an experiment. We predict here the geometric structure and pattern of spin currents in such a vortex state.

Spinful bosons can also form magnetically ordered phases with a vector order parameter. Such phases with spontaneous magnetization are indeed found in the general phase diagrams of Rashba spin-orbit-coupled bosons. However, they are conventional in the sense that they carry no spin currents that gain Rashba energy. Magnetized states host different kinds of topological defects (e.g. skyrmions) than the ones we consider here. Our focus are only the topological defects involving loop currents, which are encouraged by the spin-orbit coupling.

The structure of this paper is as follows. The detailed analysis of two-component condensates is presented in section \ref{secDoublet}. This is the simplest system that can experience a Rashba spin-orbit coupling, so we carefully derive all of our crucial results in its context. First, we explain the phase diagram and the prominent two types of condensates with uniform spin currents in subsection \ref{secDoubletLG}. Then, vortex excitations of the two types of condensates are analyzed separately in sections \ref{secDoubletVorticesT1} and \ref{secDoubletVorticesT2}. Section \ref{secTriplet} presents the similar analysis in triplet condensates, starting from the phase diagram (section \ref{secTripletLG}) and then scrutinizing the type-I and type-II vortices (sections \ref{secTripletVorticesT1} and \ref{secTripletVorticesT2}). Most findings are the same as for the two-component bosons, so we mainly focus on the differences. The section \ref{secExtraSing} briefly discusses high-energy singularities that do not exist in two-component condensates. The summary of all results, including the classification of vortex excitations and their properties, is given in the concluding section \ref{secConclusions}. The appendix \ref{ConsLaws} presents a technical derivation and analysis of current conservation laws, which support our qualitative findings from the main text, and provide a theoretical framework for calculating the detailed current patterns in the vicinity of singularities.

\section{$S=\frac{1}{2}$ condensates}\label{secDoublet}

In this section we consider a two-component boson system that can be realized in experiments with ultra-cold atoms. The two species of bosons are different hyperfine states of some bosonic element. Their energy difference can be neglected, and dynamics controlled by coupling to an external electromagnetic field. Specifically, if the two hyperfine-split internal states are interpreted as the $S^z=\pm\frac{1}{2}$ spin projections of a spin $S=\frac{1}{2}$ boson, then it may be possible in the near future to subject this artificial spin to an artificial SU(2) gauge field that implements the Rashba spin-orbit coupling in a two-dimensional confinement.

We will start with a review of the two-dimensional condensates of $S=\frac{1}{2}$ bosons with strong Rashba spin-orbit coupling from the Landau-Ginzburg perspective (section \ref{secDoubletLG}). We will find two prominent finite-momentum condensates and explore in detail their vortex excitations (sections \ref{secDoubletVorticesT1} and \ref{secDoubletVorticesT1}). These vortices can form clusters of different structure at large and short length-scales, including entire vortex lattices. Later, we will adapt this analysis to $S=1$ bosons (section \ref{secTriplet}).

\subsection{Plain $S=\frac{1}{2}$ condensates}\label{secDoubletLG}

The generic two-dimensional Landau-Ginzburg action featuring time-reversal (TR) symmetry that describes the $S=\frac{1}{2}$ bosons with a spin-orbit coupling is:
\begin{eqnarray}\label{LG2}
S_{\textrm{d}} \!\! &=& \!\! \int \dd^2 r \, \Biggl\lbrack
      \frac{1}{2m} \Bigl\lbrack\left(\boldsymbol{\nabla}-i\boldsymbol{\mathcal{A}}\right)\eta\Bigr\rbrack^{\dagger}
         \Bigl\lbrack\left(\boldsymbol{\nabla}-i\boldsymbol{\mathcal{A}}\right)\eta\Bigr\rbrack \nonumber \\
  && +t\eta^{\dagger}\eta +u(\eta^{\dagger}\eta)^{2} +b(\eta^{\dagger}\Phi_{0}\eta)^{2} \Biggr\rbrack \ .
\end{eqnarray}
The Rashba spin-orbit coupling of strength $v$ is implemented by the static SU(2) gauge field
\begin{equation}\label{GaugeField}
\boldsymbol{\mathcal{A}} = -mv(\hat{{\bf z}}\times{\bf S})
\end{equation}
whose Yang-Mills flux defined in the 2+1D space-time has only a ``magnetic field'' component $\Phi = (mv)^2$:
\begin{equation}\label{Flux}
\Phi_0 = \epsilon^{0\mu\nu} \Bigl( \partial_\mu \mathcal{A}_\nu - i \mathcal{A}_\mu \mathcal{A}_\nu \Bigr) = \Phi S^z
\end{equation}
($\epsilon^{\mu\nu\lambda}$ is the Levi-Civita tensor). Both the gauge field and flux components (temporal $\mu=0$ and spatial $\mu\in\lbrace x,y \rbrace$) are SU(2) matrices expressed in terms of the spin operators ${\bf S} = (S^x, S^y, S^z)$. Presently, we are dealing with an $S=\frac{1}{2}$ representation, so we may choose ${\bf S} = \frac{1}{2} \boldsymbol{\sigma}$, where $\boldsymbol{\sigma} = (\sigma^x, \sigma^y, \sigma^z)$ are Pauli matrices.

The most general two-component order parameter
\begin{equation}\label{OP2}
\eta = \left(\begin{array}{c} \eta_{\uparrow} \\ \eta_{\downarrow} \end{array}\right)
  = \zeta\left(\begin{array}{c} \cos\left(\frac{\alpha}{2}\right)\, e^{-i\frac{\theta}{2}} \\ i\sin\left(\frac{\alpha}{2}\right)\, e^{i\frac{\theta}{2}}
    \end{array}\right)e^{i\gamma} 
\end{equation}
is specified in terms of one real amplitude $\zeta$ and three angles $\alpha,\theta,\gamma$. We will characterize various condensates by the density $j_0^{\phantom{a}}$ and current ${\bf j}$ of ``charge'', as well as the densities $J_0^a$ and currents ${\bf J}^a$ of all spin projections $a\in\lbrace x,y,z \rbrace$:
\begin{eqnarray}\label{Currents}
j_{0} = \eta^{\dagger}\eta &~,~&
{\bf j} = -\frac{i}{2m}\Bigl(\eta^{\dagger}(\boldsymbol{\nabla}\eta)-(\boldsymbol{\nabla}\eta^{\dagger})\eta\Bigr) \\
J_{0}^{a} = \eta^{\dagger}S^{a}\eta\nonumber &~,~&
{\bf J}^{a} = -\frac{i}{2m}\Bigl(\eta^{\dagger}S^{a}(\boldsymbol{\nabla}\eta)-(\boldsymbol{\nabla}\eta^{\dagger})S^{a}\eta\Bigr)
 \ . \nonumber
\end{eqnarray}
Note that these are not the conserved SU(2) gauge-covariant currents. Unless we emphasize otherwise, we will refer to (\ref{Currents}) simply as currents throughout the paper. They are measurable because the SU(2) gauge symmetry is explicitly broken in cold atom and solid state systems that exhibit the Rashba spin-orbit coupling. In a true SU(2) gauge-invariant system with dynamical SU(2) gauge fields, only the gauge-covariant currents would be measurable; we will need them for the analysis of conservation laws in appendix \ref{ConsLaws}.

The Rashba spin-orbit coupling contributes
\begin{equation}\label{RashbaEnergy}
\mathcal{E}_{\textrm{R}} = mv\left(J_{x}^{y}-J_{y}^{x}\right)
\end{equation}
to the energy density in terms of the spin currents, so it favors a ``helical'' flow of spin: the orientation of the flowing spin is perpendicular to the flow direction according to the right-hand rule. There are two characteristic ways to obtain such spin currents. First, a spin current can be established without inducing any uniform charge current or spin polarization. We will classify such states with pure spin currents as ``type-I''. Alternatively, a spin current can result from the movement of particles that have a definite spin. The ensuing condensates, that we will denote as ``type-II'', also carry a uniform charge current and have a uniform component of spin polarization. Note that no two-component condensate can have all local spin projections equal to zero, so the artificial TR symmetry is automatically broken in any superfluid state. However, the type-I condensates of $S=1$ bosons can be TR-invariant.

Both type-I and type-II states can exist in equilibrium despite carrying finite currents (\ref{Currents}). This is the consequence of the background SU(2) gauge field that alters the conservation laws. The conserved charge current is even finite and periodically modulated in the $S=\frac{1}{2}$ type-I state \cite{Ozawa2012a}, but this nuisance is eliminated in the $S=1$ condensate. We will not pursue these details, but rather focus on generic qualitative properties of the condensates in order to later analyze the character of their vortex excitations.

\begin{figure}
\includegraphics[width=3.0in]{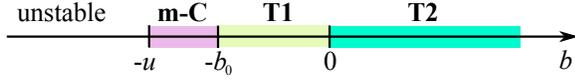}
\caption{\label{LG2pd}The phase diagram of superfluid states of two-component bosons with Rashba spin-orbit coupling. m-C is a conventional uniform condensate with spin-polarization along the $z$-axis. The type-I or ``striped'' condensate (T1), and type-II or ``plain wave'' condensate (T2) carry helical spin currents.}
\end{figure}

The mean-field phase diagram of superfluid phases is summarized in Fig.\ref{LG2pd}, assuming that the coupling $t$ is small or negative. Normal state is established for sufficiently large $t>0$. This system is simple enough to determine the phase diagram analytically, and we present the detailed derivation in the remainder of this subsection. A reader not interested in these details can safely skip to the next subsection.

The currents (\ref{Currents}) can be readily expressed in terms of the four parameters $(\zeta, \alpha, \theta, \gamma)$ to characterize any condensate:
\begin{eqnarray}\label{CurDen1}
j_{0} &=& \zeta^{2} \\
J_{0}^{x} &=& -\frac{\zeta^{2}}{2}\sin\alpha\sin\theta \nonumber \\
J_{0}^{y} &=& \frac{\zeta^{2}}{2}\sin\alpha\cos\theta \nonumber \\
J_{0}^{z} &=& \frac{\zeta^{2}}{2}\cos\alpha \nonumber \ , \nonumber
\end{eqnarray}
\begin{eqnarray}\label{CurDen2}
{\bf j} &=& \frac{\zeta^{2}}{m}\left\lbrack \boldsymbol{\nabla}\gamma-\frac{1}{2}\cos\alpha\boldsymbol{\nabla}\theta\right\rbrack \\
{\bf J}^{x} &=& \frac{\zeta^{2}}{2m}\left\lbrack \frac{1}{2}\cos\theta\boldsymbol{\nabla}\alpha
  -\sin\alpha\sin\theta\boldsymbol{\nabla}\gamma \right\rbrack \nonumber \\
{\bf J}^{y} &=& \frac{\zeta^{2}}{2m}\left\lbrack \frac{1}{2}\sin\theta\boldsymbol{\nabla}\alpha
  +\sin\alpha\cos\theta\boldsymbol{\nabla}\gamma \right\rbrack \nonumber \\
{\bf J}^{z} &=& \frac{\zeta^{2}}{2m}\left\lbrack \cos\alpha\boldsymbol{\nabla}\gamma-\frac{1}{2}\boldsymbol{\nabla}\theta\right\rbrack
  \nonumber \ .
\end{eqnarray}
The condensates with uniform helical currents can be parametrized by six scalars $(\zeta,\alpha_{0},\theta_{0},k_{\alpha},k_{\theta},k_{\gamma})$ and
\begin{equation}
\alpha=\alpha_{0}+{\bf k}_{\alpha}{\bf r}\quad,\quad\theta=\theta_{0}+{\bf k}_{\theta}{\bf r}\quad,\quad\gamma={\bf k}_{\gamma}{\bf r}
\end{equation}
combined with $\zeta=\textrm{const}$. The energy density in (\ref{LG2}) expressed in terms of these parameters is:
\begin{eqnarray}\label{LG2b}
\mathcal{E} &=& \frac{1}{2m}\left\lbrace \zeta^{2}\left\lbrack k_{\gamma}^{2}+\frac{1}{4}k_{\theta}^{2}+\frac{1}{4}k_{\alpha}^{2}
                -\cos\alpha\,{\bf k}_{\theta}{\bf k}_{\gamma}\right\rbrack \right\rbrace \nonumber \\
  && +\frac{v\zeta^{2}}{4}\left\lbrack \frac{1}{2}(\hat{\bf z}\times\hat{\boldsymbol{\theta}}){\bf k}_\alpha
     -\sin\alpha\,\hat{\boldsymbol{\theta}}{\bf k}_\gamma \right\rbrack \nonumber \\
  && +t'\zeta^{2}+u\zeta^{4}+b\zeta^{4}\cos^{2}\alpha \ ,
\end{eqnarray}
where we defined $t'=t+\frac{1}{4}mv^2$ and used
\begin{equation}\label{ThetaHat0}
\hat{\boldsymbol{\theta}}=\hat{{\bf x}}\cos\theta+\hat{{\bf y}}\sin\theta \ .
\end{equation}

We now wish to minimize this energy. It is apparent that the spatial oscillations of $\theta$ cannot help lower the energy: they cost kinetic energy, while bringing the average Rashba energy gain ($\propto v$) to zero. Hence, we should set ${\bf k}_{\theta}=0$. The only way to perhaps gain some Rashba energy from the oscillations of $\theta$ is to correlate them with the oscillations of $\alpha$ and make $\langle\sin\alpha\sin\theta\rangle\neq0$ on average. However, this would also require ${\bf k}_{\gamma}\neq0$ and hence cost more kinetic energy than other states.

The prominent states with ${\bf k}_{\theta}=0$ can be classified by whether $k_\alpha$ is zero or finite. If $k_\alpha\neq 0$, then $\sin\alpha$ and $\cos\alpha$ average out to zero, while $\sin^2\alpha$ and $\cos^2\alpha$ average out to $\frac{1}{2}$. The average energy density in this case
\begin{eqnarray}\label{EDen1}
\mathcal{E}({\bf k}_{\alpha}\neq0) &=& \frac{\zeta^{2}}{2m}\left(k_{\gamma}^{2}+\frac{1}{4}k_{\alpha}^{2}\right)
   +\frac{v\zeta^{2}}{8}(\hat{\bf z}\times\hat{\boldsymbol{\theta}})\,{\bf k}_\alpha \nonumber \\
&& +t'\zeta^{2}+\left(u+\frac{b}{2}\right)\zeta^{4}
\end{eqnarray}
is clearly minimized when ${\bf k}_\gamma=0$ and $\hat{\boldsymbol{\theta}}$ is perpendicular to ${\bf k}_\alpha$. According to (\ref{CurDen1}) and (\ref{CurDen2}), this corresponds to a helical spin current (spin projection being perpendicular to its flow direction). There is no flow of charge (${\bf j}=0$), so we classify this state as type-I. It can be easily seen that the optimum $|{\bf k}_\alpha| = \frac{1}{2}mv$ yields the minimum energy density $\mathcal{E}_{\textrm{I}} \to -t''^2/4u'$ in a superfluid state, where $t''=t+\frac{7}{32}mv^2<0$ and $u'=u+\frac{b}{2}$. Alternatively,
\begin{eqnarray}\label{EDen2}
\mathcal{E}({\bf k}_{\alpha}=0) &=& \frac{\zeta^{2}}{2m}k_{\gamma}^{2}
   -\frac{v\zeta^{2}}{4}\sin\alpha_{0}\,\hat{\boldsymbol{\theta}}{\bf k}_\gamma \nonumber \\
&& +t'\zeta^{2}+\left(u+b\cos^{2}\alpha_{0}\right)\zeta^{4} \ .
\end{eqnarray}
is minimized through the Rashba term when $\alpha_0=\frac{\pi}{2}$ and $\hat{\boldsymbol{\theta}} \parallel {\bf k}_\gamma $. This time, the ensuing helical spin current is created by the flow of charge in a spin-polarized background, so we classify it as a type-II condensate. The optimum $|{\bf k}_\gamma| = \frac{1}{4}mv$ enables this energy density to reach the smallest value $\mathcal{E}_{\textrm{II}} \to -t''/4u$, where $t''$ is the same as above, but the quartic coupling is just the original $u$.

It is now clear that $\mathcal{E}_{\textrm{I}} < \mathcal{E}_{\textrm{II}}$ if $u'<u$, so that type-I condensates can be stable only when $b<0$, and type-II condensates are found for $b>0$. However, the energy density (\ref{EDen2}) can be better than (\ref{EDen1}) even for $b<0$ if a large value of $|b|$ selects $\alpha_0 \in \lbrace 0,\pi \rbrace$. In order for this to happen, the resulting state must be fully spin-polarized in the $z$-direction, and gain nothing from the Rashba spin-orbit coupling. It should then minimize its kinetic energy by carrying no currents. The lowest energy density it can reach is $\mathcal{E}_{\textrm{m}} \to -t'^2/4(u+b)$. This surpasses $\mathcal{E}_{\textrm{I}}$ when $-u<b<-b_0<0$, where $b_0/u = 1-t'^2/(2t''^2-t'^2)$. The resulting complete phase diagram of superfluid states (when they have lower energy than the normal state) is shown in Fig.\ref{LG2pd}.

\subsection{Type-I vortices}\label{secDoubletVorticesT1}

The following analysis is devoted to vortices in type-I condensates. Our goal is to describe the patters of spin currents surrounding a singularity, reveal the quantum numbers carried by vortices, and establish the fundamental properties of vortex clusters. Each of the three angles $\alpha, \theta, \gamma$ in the order parameter (\ref{OP2}) can have topological defects. Therefore, the dynamics features U(1) singularities of ``helical'' ${\bf J}^{x,y} \sim \boldsymbol{\nabla}\alpha$ and ``chiral'' ${\bf J}^z \sim \boldsymbol{\nabla}\theta$ spin-currents, as well as charge ${\bf j} \sim \boldsymbol{\nabla}\gamma$ currents. This classification of vortices based on the fundamental types of currents is convenient and motivated by the symmetry of the spin-orbit flux (\ref{Flux}), even though the gauge field (\ref{GaugeField}) does not conserve any spin projection and yields intricate correlations between the singular structures of different order parameter components.

We will identify different possibilities for vortex structures at large and short length-scales. The helical and pure charge vortices interact via the usual Coulomb potential at large distances (logarithmic function of distance in two-dimensions). The chiral vortices lower their energy by binding to charge vortices, but unavoidably cost a lot and interact among themselves via a linear potential like quarks in quantum chromodynamics. The simplest large neutral clusters of vortices are dipoles. Additional singular structures are possible at short length-scales comparable to the SU(2) ``magnetic length'' of the Rashba spin-orbit coupling. Small metastable quadruplets of chiral vortices can be either created as isolated excitations, or tiled into a dense (meta)stable vortex lattice. We will derive these qualitative conclusions merely from energy considerations here, and support them by current conservation laws in appendix \ref{ConsLaws}.

In order to study the lowest energy vortex excitations, we will conveniently rewrite the order parameter (\ref{OP2}) in a slightly different form:
\begin{equation}\label{OP2b}
\eta = \zeta\left(\begin{array}{c}
  \cos\left(\frac{\alpha}{2}\right)\, e^{i\theta_{\uparrow}}\\
  i \sin\left(\frac{\alpha}{2}\right)\, e^{i\theta_{\downarrow}}
\end{array}\right) \ .
\end{equation}
This allows us to transparently represent the binding of chiral and charge vortices, where only one of the two angles $\theta_\uparrow, \theta_\downarrow$ winds by $2\pi$. Such binding saves energy because the amplitude of only one order-parameter component needs to be depleted near the vortex core. A pure chiral vortex without any charge currents is actually a pair of a coinciding  $\theta_\uparrow$ vortex and a $\theta_\downarrow$ antivortex, leading to the total winding of the chiral angle $\theta = \theta_\downarrow - \theta_\uparrow$ by $4\pi$. Pure charge vortices in the absence of circulating spin-currents are more easily represented by (\ref{OP2}) with a constant $\theta$ and winding $\gamma = \frac{1}{2}(\theta_\downarrow + \theta_\uparrow)$.

Spin currents are determined by the gradient energy extracted from (\ref{LG2}):
\begin{eqnarray}\label{Gr1}
E_{\textrm{kin}} \!\!&=&\!\! \int \dd^2 r \; \Bigl\lbrace \frac{1}{2m} |\boldsymbol{\nabla} \eta|^2
   + mv \left( J_y^x - J_x^y \right) \Bigr\rbrace \\
\!\!&\propto&\!\! \int \dd^2 r \; \biggl\lbrace
  (\boldsymbol{\nabla}\zeta)^{2} + \frac{\zeta^2}{4}(\boldsymbol{\nabla}\alpha)^{2} \nonumber \\
&& + \frac{\zeta^{2}}{2} \Bigl\lbrack (1+\cos\alpha)(\boldsymbol{\nabla}\theta_{\uparrow})^{2}
    +(1-\cos\alpha)(\boldsymbol{\nabla}\theta_{\downarrow})^{2}
  \Bigr\rbrack \nonumber \\
&& +\frac{mv\zeta^{2}}{4} \biggl\lbrack
     (\hat{{\bf z}}\times\hat{\boldsymbol{\theta}})\boldsymbol{\nabla}\alpha
    -\sin\alpha\,\hat{\boldsymbol{\theta}}(\boldsymbol{\nabla}\theta_{\uparrow}+\boldsymbol{\nabla}\theta_{\downarrow})
   \biggr\rbrack \biggr\rbrace \nonumber
\end{eqnarray}
The quadratic gradient terms are pure kinetic energy and always positive. However, the Rashba energy (proportional to $v$) is linear in gradients, so it can be negative and stabilize a state of non-zero currents. The unit-vector
\begin{equation}\label{ThetaHat}
\hat{\boldsymbol{\theta}} = \hat{\bf x}\cos(\theta_\downarrow-\theta_\uparrow) + \hat{\bf y}\sin(\theta_\downarrow-\theta_\uparrow) 
\end{equation}
represents the direction of spin that is transported by the helical current ${\bf J}^{x,y} \sim {\boldsymbol{\nabla}\alpha}$ in a type-I state.

\subsubsection{Large-scale vortex structure}\label{T1LargeScale}

The helical angle $\alpha$ completes one period of oscillations across the distance $(mv)^{-1}$, which can be very small if the spin-orbit coupling $v$ is strong, or mass $m$ large. Such rapid oscillates can be coarse-grained to reveal the structure of spin-currents at large length-scales. The coarse-grained gradient energy (\ref{Gr1}) is:
\begin{eqnarray}\label{Gr2}
E_{\textrm{kin}} \!\!&\propto&\!\! \int \dd^2 r \; \biggl\lbrace (\boldsymbol{\nabla}\zeta)^{2}
  +\frac{\zeta^{2}}{2}\Bigl\lbrack(\boldsymbol{\nabla}\theta_{\uparrow})^{2}+(\boldsymbol{\nabla}\theta_{\downarrow})^{2}\Bigr\rbrack
  \nonumber \\
&& +\frac{\zeta^{2}}{4}\left(\boldsymbol{\nabla}\alpha+\frac{mv}{2}\hat{\bf z}\times\hat{\boldsymbol{\theta}}\right)^{2} \biggr\rbrace
\end{eqnarray}
up to a correction to the chemical potential.

Helical vortices are the simplest ones to understand. A helical vortex perturbs the uniform helical current $\boldsymbol{\nabla}\alpha$ of the type-I condensate by a U(1) singularity in the phase $\alpha$, placed, for instance, at the origin:
\begin{equation}
\alpha({\bf r}) = {\bf kr} + q \phi \ .
\end{equation}
Here we expressed the coordinates ${\bf r}=(r,\phi)$ in the cylindrical system. The helical vorticity must be quantized, $q \in \mathbb{Z}$. The phases $\theta_\uparrow$ and $\theta_\downarrow$ are to remain mostly uniform and unaffected by the singularity. If we optimally set ${\bf k} = -\frac{mv}{2} \, \hat{\bf z}\times \hat{\boldsymbol{\theta}}$ according to the bulk orientation of $\hat{\boldsymbol{\theta}}$, then the kinetic and Rashba energy of this configuration is:
\begin{eqnarray}\label{Gr3}
E_{\textrm{kin}} \!\!&\propto&\!\! \int \dd^2 r \biggl\lbrace (\boldsymbol{\nabla}\zeta)^{2}
  +\zeta^{2} \Bigl( \boldsymbol{\nabla}\alpha + \frac{mv}{2} \, \hat{\bf z} \times \hat{\boldsymbol{\theta}} \Bigr)^2 \biggr\rbrace
  \nonumber \\
\!\!&=&\!\! \int \dd^2 r \biggl\lbrace (\boldsymbol{\nabla}\zeta)^{2} + \frac{1}{r^2}\zeta^{2} \biggr\rbrace \ ,
\end{eqnarray}
which scales as $\log(R)$ with the system size $R$. The infra-red divergence of this energy can be cut off by another vortex of charge $-q$ placed a distance $l$ away from the $q$ vortex. The ensuing potential energy of inter-vortex interactions is a logarithmic function of $l$. As in the ordinary superfluids, two helical vortices of charge $q_1$ and $q_2$ will interact via the potential:
\begin{equation}
V_\alpha (l) = -4\pi K \, q_1 q_2 \, \log\left(\frac{l}{\xi}\right) \ ,
\end{equation}
where $K\propto \zeta^2$ is the superfluid stiffness, and $\xi$ is the vortex core size.

\begin{figure}
\includegraphics[width=2.5in]{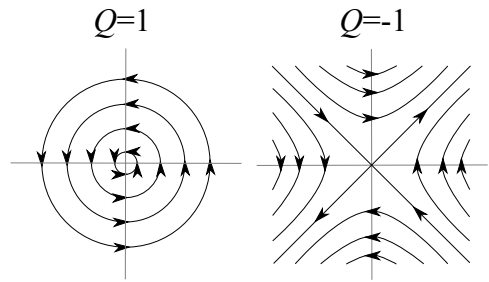}
\caption{\label{VecVortex}Topological defects of two-dimensional vector fields ${\bf v} \propto -\hat{\bf x}\sin(Q\phi) + \hat{\bf y}\cos(Q\phi)$, where $\phi$ is the polar angle, can implement any total rotation angle $2\pi Q$ of the vector ${\bf v}$ around a loop that encloses the singularity. However, only the $Q=1$ case can describe a current field of a quantized U(1) vortex, ${\bf v} = \boldsymbol{\nabla}\theta$. Any singular configuration of $\theta = k\phi$, which winds by $2\pi k$, $k\neq 0$ on the loop around the singularity, still corresponds to the vector field ${\bf v} = k(-\hat{\bf x}\sin\phi + \hat{\bf y}\cos\phi)$ of a $Q=1$ vector-vortex. Therefore, the curl of ${\bf v}$ can be zero only if $Q=0$ or $Q=1$.}
\end{figure}

It turns out that the topological defects formed by the circulating ${\bf J}^z \sim \boldsymbol{\nabla}\theta$ spin-currents are more complicated. As noted before, a chiral vortex binds to itself a charge vortex in order to minimize energy, producing a U(1) singularity of only one of the two angles $\theta_\uparrow, \theta_\downarrow$. Regardless of which angle ends up winding, only the pure chiral angle $\theta = \theta_\downarrow - \theta_\uparrow$ is coupled to the helical currents $\boldsymbol{\nabla}\alpha$ in (\ref{Gr2}) and winds by $\pm 2\pi$. Let us, then, consider a generalized single chiral vortex of charge $Q\neq 0$, where the total chiral angle $\theta$ winds by $2\pi Q$. The vector $\hat{\boldsymbol{\theta}}$ in (\ref{ThetaHat}) rotates by the quantized angle
\begin{equation}
\oint \dd {\bf l} \boldsymbol{\nabla}\theta = 2\pi Q\quad,\quad Q \in \mathbb{Z}
\end{equation}
on any loop that encloses the singularity. Chiral spin-currents unavoidably contribute a kinetic energy that diverges logarithmically with the system size. We can at least try to keep the total cost of helical currents finite by maintaining the optimum condition for type-I patterns
\begin{equation}\label{HelicalT1}
\boldsymbol{\nabla} \alpha = -\frac{mv}{2} \hat{\bf z} \times \hat{\boldsymbol{\theta}}
\end{equation}
everywhere in space. This implies that the rotation of $\hat{\boldsymbol{\theta}}$ on loops must be matched by the equivalent rotation of $\boldsymbol{\nabla}\alpha$. Therefore, the resulting vector field of $\boldsymbol{\nabla}\alpha$ should ideally have a topological defect (vector-vortex) of the same quantized charge $Q$. Examples of vector-vortices are shown in Fig.\ref{VecVortex}. 

If a chiral singularity sits at the origin, and $\theta$ winds by $2\pi Q$ on loops around the origin, then $\theta(r,\phi)=Q\phi - \theta_0$ expressed in cylindrical coordinates is the configuration that minimizes the $(\boldsymbol{\nabla}\theta)^2$ energy of chiral spin-currents in (\ref{Gr3}). We could try to simultaneously minimize the kinetic energy of helical spin-currents $(\boldsymbol{\nabla}\alpha - \boldsymbol{\delta\alpha})^2$, where
\begin{eqnarray}\label{Da}
\boldsymbol{\delta\alpha} \!\!&=&\!\! -\frac{mv}{2}(\hat{{\bf z}}\times\hat{\boldsymbol{\theta}}) \\
  \!\!\!\!&=&\!\! \frac{mv}{2}\Bigl\lbrack \hat{{\bf r}}\sin\Bigl((Q-1)\phi-\theta_0\Bigr)
              -\hat{\boldsymbol{\phi}}\cos\Bigl((Q-1)\phi-\theta_0\Bigr)\Bigr\rbrack \nonumber
\end{eqnarray}
The curl of $\boldsymbol{\delta\alpha}$
\begin{equation}\label{AlphaCurl}
\hat{{\bf z}}(\boldsymbol{\nabla}\times\boldsymbol{\delta a})=-\frac{mv}{2r}Q\cos\Bigl((Q-1)\phi-\theta_{0}\Bigr)
\end{equation}
vanishes if $Q=0$, or $Q=1$ and $\theta_0 = \pm\pi/2$. Therefore, the optimal pattern of helical spin currents $\boldsymbol{\nabla}\alpha \to \boldsymbol{\delta\alpha}$ is indeed given by (\ref{Da}) in the vicinity of a $Q=+1$ chiral vortex. However, the above curl does not vanish in the case of any finite $Q\neq 1$, thus making $\boldsymbol{\delta\alpha}$ different from the gradient of any scalar. The best we can do in such cases is separate $\boldsymbol{\delta\alpha} = \boldsymbol{\nabla}\alpha + \hat{\bf z}\times\boldsymbol{\nabla}\alpha'$ into its pure gradient and ``curl'' parts, for which we need two scalars $\alpha$ and $\alpha'$ respectively in two spatial dimensions. If (\ref{AlphaCurl}) does not vanish, we can readily find:
\begin{equation}
\alpha'(r,\phi) = \frac{mvr}{2} f(\phi) \ ,
\end{equation}
where
\begin{equation}
f(\phi) = \left\lbrace \begin{array}{ccc}
    \frac{1}{2-Q}\cos\Bigl((Q-1)\phi-\theta_{0}\Bigr) &,& Q\neq 2 \\
    \phi\sin\Bigl(\phi-\theta_{0}\Bigr) &,& Q=2 \end{array} \right\rbrace \ .
\end{equation}
The energy cost (\ref{Gr3}) cannot be reduced below that determined by the ``curl'' part:
\begin{eqnarray}
E_{\textrm{kin}} \!\!&\propto&\!\! \int d^{2}r\,\zeta^{2}\Bigl(\boldsymbol{\nabla}\alpha-\boldsymbol{\delta a}\Bigr)^{2}\!\!+\cdots
  > \int d^{2}r\,\zeta^{2}(\hat{{\bf z}}\times\boldsymbol{\nabla}\alpha')^{2} \nonumber \\
  \!\!&=&\!\! \int d^{2}r\,\left(\frac{mv\zeta}{2}\right)^2\Bigl\lbrack f^2(\phi)+f'^2(\phi) \Bigr\rbrack \ .
\end{eqnarray}
This scales as the system area $R^2$, and by far exceeds the $\log(R)$ energy of a $Q=+1$ vortex for which we can simply choose $\alpha'=0$.

Now we face a problem. A $Q=+1$ vortex, despite being relatively cheap, cannot be excited in an infinite system without a compensation by a $Q=-1$ vortex. However, it seems that a $Q=-1$ vortex costs forbiddingly high energy. We have no reason to expect much symmetry between the $Q=+1$ and $Q=-1$ singularities, since they are not related to each other by the TR. But surely we must do better and construct a different pattern of vortex spin-currents that costs less energy.

\begin{figure}
\subfigure[{}]{\includegraphics[height=1.8in]{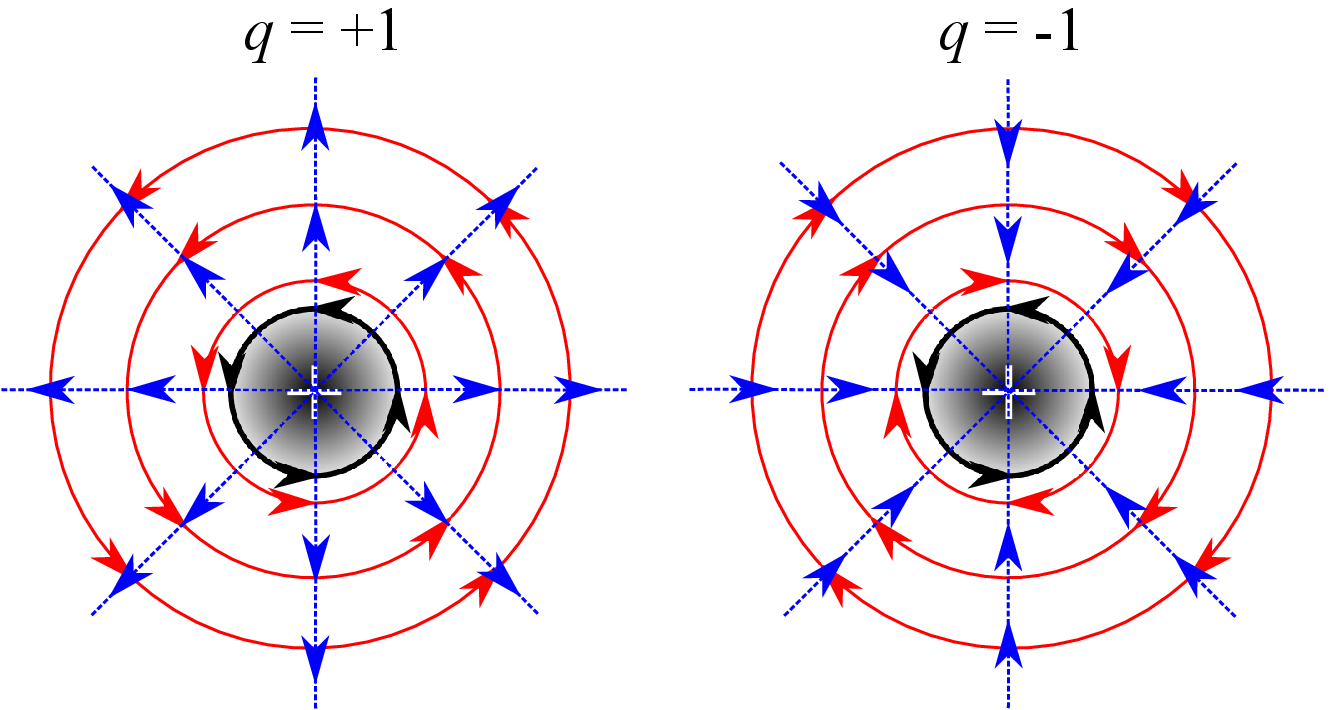}}
\subfigure[{}]{\includegraphics[height=1.47in]{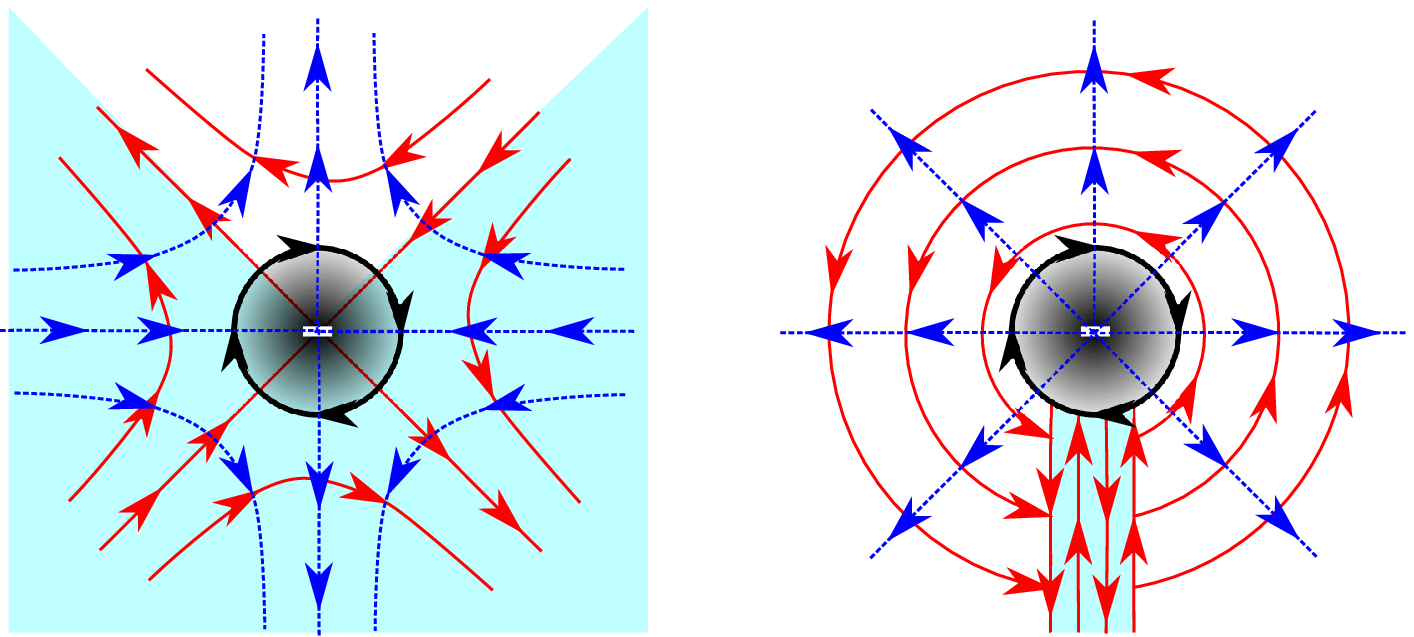}}
\caption{\label{VorT1}(a) The structure of chiral $Q=Q'=1$ vortices. The shaded circle at the center is a vortex core, and the solid oriented line at the core border shows the flow of chiral spin-currents ${\bf J}^z \sim \boldsymbol{\nabla}\theta$. The solid red lines indicate $\hat{\boldsymbol{\theta}}$, which is the \emph{orientation of the spin component} that flows in a helical spin-current. Dashed blue lines show the \emph{flow direction} of the helical current ${\bf J}^{x,y} \sim \boldsymbol{\nabla}\alpha$. (b) The structure of $Q=-1$, $Q'=1$ vortices and the emergence of strings. The left panel shows the naive patterns of \emph{non-quantized} helical spin-currents ($\boldsymbol{\delta\alpha}$) based on the simple $\theta$-winding and the requirement (\ref{HelicalT1}). The right panel visualizes the formation of a string that keeps the same $\theta$ winding while fixing the quantization of helical currents (by removing their curls from the bulk). The entire shaded region is compressed into a filament.}
\end{figure}

A strong spin-orbit coupling will rearrange the configuration of $\theta$ in such a way that the condition (\ref{HelicalT1}) can be satisfied almost everywhere in space. According to (\ref{Da}) and (\ref{AlphaCurl}), this is possible only for a purely uniform ($Q\to 0$), or a radial ($Q\to +1$) local arrangement of $\boldsymbol{\nabla}\alpha$ near a singularity. Therefore, the bulk of the space surrounding any chiral singularity must look like a $Q'=0$ or $Q'=1$ vortex. If $\theta$ is to wind by $2\pi Q$, $Q\neq Q'$ on a loop that encloses the singularity, then a string-shaped area of compressed $\theta$ and $\alpha$ oscillations must emanate from the singularity, and $\theta$ must unwind by $2\pi(Q-Q')$ across it. The kinetic energy of the bulk oscillations of $\alpha$ and $\theta$ now scales as $\log(R)$ with the system radius $R$ in the worst case, but the string costs energy proportional to its length $R$. This is still a large energy, but much smaller than $R^2$ in the limit $R\to\infty$ of a large system size. Fig.\ref{VorT1} illustrates the emergence of strings.

We now have a good chance to enable compensation of the chiral vortex charges, provided that strings can be finite and terminated at both ends by singularities. This restricts the apparent ``bulk'' winding numbers $Q'$ of singularities. Consider a collection of $N$ singularities, with total charges $Q_i$ and ``bulk'' charges $Q'_i$, $i=1,\dots,N$. There could be $n_i$ strings attached to the $i^\textrm{th}$ vortex, each bringing some amount of chiral charge $\Delta Q_{i,k}$, $k=1,\dots,n_i$ to the vortex. Strings have an inherent orientation, due to the definite sense of $\theta$ winding across them. Therefore, a string that connects vortices $i$ and $j$ can bring to the vortex $i$ only the amount of chiral charge that it took from the vortex $j$. The total $\theta$ winding across the ``bulk'' ($Q'_i$) and all the strings ($\sum \Delta Q_{i,k}$) must add up to the total winding ($Q_i$) for any vortex $i$:
\begin{equation}
Q'_i + \sum_{k=1}^{n_i} \Delta Q_{i,k} = Q_i \ .
\end{equation}
If we now add these equations for all vortices, we find:
\begin{equation}\label{Qp}
\sum_{i=1}^N Q'_i = \sum_{i=1}^N Q_i - \sum_{i=1}^N \sum_{k=1}^{n_i} \Delta Q_{i,k} = 0 \ .
\end{equation}
The first term on the right-hand side is zero because the cluster of vortices must be neutral, while the second is zero because strings themselves are not sources of chiral charge. Since the only allowed values for $Q'$ are $0$ and $1$, we conclude that all vortices of a neutral cluster must have $Q'=0$. The winding of $\theta$ is completely consumed by the strings, so $\theta$ remains overall uniform in the bulk. The ensuing configuration of $\boldsymbol{\nabla}\alpha$ in the bulk is also uniform. The inter-vortex interaction potential is linear in the distance between vortices, or more accurately, proportional to the string length. Chiral vortices very much resemble quarks in quantum chromodynamics, and experience confinement and asymptotic freedom.

The only way to avoid the formation of strings while having non-zero winding of $\theta_\uparrow$ or $\theta_\downarrow$ is to eliminate the singular rotations of $\hat{\boldsymbol{\theta}}$. This necessitates $\theta_\uparrow = \theta_\downarrow + \textrm{const}$ and thus corresponds to a pure charge vortex without any circulating spin-currents. Since $\hat{\boldsymbol{\theta}}$ is held fixed, helical spin-currents remain uniform as in the bulk type-I state, and charge vortices only pay the price of extra charge currents. The energy cost of an uncompensated charge vortex is logarithmically divergent with system size, and the potential energy of interactions between them is a logarithmic function of their distance.

\subsubsection{Short-scale vortex structure}

The physical picture of chiral vortices developed in the previous section holds primarily at length-scales large in comparison to $(mv)^{-1}$. The strings emanating from such singularities are fully formed only at distances much larger than $(mv)^{-1}$ from the vortex core, and their thickness is at best comparable to $(mv)^{-1}$. Coarse-grained spin-current conservation laws that we will formulate later support these conclusions. There is still room for the existence of other vortex structures, as long as they are neutralized at length-scales of the order of $(mv)^{-1}$. We will analyze them primarily by following the optimal condition (\ref{HelicalT1}), which holds microscopically and independently of coarse-graining according to (\ref{Gr1}).

We discovered in (\ref{AlphaCurl}) that an isolated $Q=+1$ chiral vortex need not be a terminal point of a string. In that case, shown in Fig.\ref{VorT1}(a), its energy cost scales only logarithmically with the system size. Its ``bulk'' charge $Q'=1$ supports a radially-aligned arrangement of helical spin-currents. Maintaining the optimal value $|\boldsymbol{\nabla}\alpha| = \frac{mv}{2}$ requires a spatial distribution of sources or drains for helical currents, which interestingly do not contradict current conservation laws (see appendix \ref{ConsLaws}). We could say that a $Q=+1$ chiral vortex carries a secondary charge $\pm 1$, depending on whether the helical currents have a source or drain configuration in its vicinity. The ensuing pattern of $\boldsymbol{\nabla}\alpha$ has zero curl. Recall that a string must be attached to a $Q=+1$ vortex only if we want to cut-off its infra-red divergent energy by a distant $Q=-1$ antivortex. But the string can ``diffuse'' and open up in all directions if the two connected singularities are close enough to each other. The lowest-energy $Q=+1$ vortex then provides a source or drain for helical currents.

As soon as we compensate the chiral charge of a $Q=+1$ vortex by a nearby $Q=-1$ antivortex, $\hat{\boldsymbol{\theta}}$ becomes uniform far away from the vortex dipole and demands a uniform flow of helical currents in the bulk. However, the antivortex is not a net source or drain of helical spin-currents. Regardless of how the field lines actually arrange, the uncompensated localized helical source that surrounds the $Q=+1$ singularity yields a kinetic energy build-up that scales as $\log(R)$ with the system size $R$. This energy is analogous to the 2D Coulomb energy of a point ``charge'' that creates an ``electrostatic'' field $\boldsymbol{\nabla}\alpha$. Given the Rashba interaction between the helical current $\boldsymbol{\nabla}\alpha$ and the chiral angle $\theta$, it is tempting to deform the configuration of $\theta$ and fully compensate this Coulomb energy. However, the ensuing Coulomb-like configuration of $\boldsymbol{\nabla}\theta$ again costs a logarithmically divergent energy. Instead of eliminating helical sources by attaching strings to vortices, we can tame the divergent energy cost of a source-like dipole by introducing a new drain-like dipole. This gives rise to a neutral quadruplet of chiral vortices.

\begin{figure}
\includegraphics[height=2.5in]{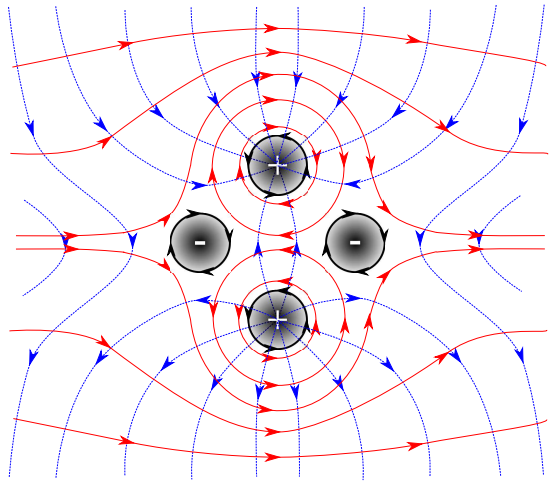}
\caption{\label{VorQuad}The simplest neutral quadruplet of vortices. Vortex cores are shaded in gray, surrounded by thick oriented black lines that depict the flow of ${\bf J}^z\sim\boldsymbol{\nabla}\theta$ spin currents. Solid thin red lines show the local orientation of the vector $\hat{\boldsymbol{\theta}}$. Dotted thin blue lines are the ``field lines'' of $\boldsymbol{\nabla}\alpha$, hence showing the flow direction of helical spin currents ${\bf J}^{x,y}$. Far away from the quadruplet, the order parameter is reduced to that of a mean-field type-I condensate.}
\end{figure}

\begin{figure}
\includegraphics[width=3.3in]{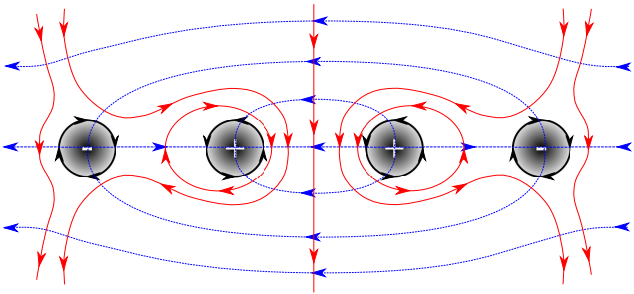}
\caption{\label{VorQuad2}An example of a quadruplet that can combine the singularities of both $\theta_\uparrow$ and $\theta_\downarrow$. Going from left to right, the winding chiral angles can be $\uparrow\downarrow\uparrow\downarrow$, $\uparrow\uparrow\downarrow\downarrow$,  and the two inverted combinations, among which the first one and its reflection have the lowest energy. The color and line convention is the same as in Fig.\ref{VorQuad} (so that the configuration of $\theta = \theta_\downarrow-\theta_\uparrow$ is shown instead of either $\theta_\downarrow$ or $\theta_\uparrow$).}
\end{figure}

Figures \ref{VorQuad} and \ref{VorQuad2} show several examples of type-I chiral quadruplets, by depicting classically desired patterns of helical currents. These patterns of helical currents are naively constructed to satisfy (\ref{HelicalT1}) for the given arrangements of chiral vortices, but unavoidably introduce vorticity of $\alpha$ near the chiral antivortices whose spatial average is zero. The actual flow of helical supercurrents must be corrected in order to quantize this $\alpha$ vorticity, but it still qualitatively resembles the shown classical configurations. The condition (\ref{HelicalT1}) cannot be perfectly satisfied, but the excess energy cost is limited by the quadruplet's finite area. The ensuing vortex cluster has both chiral and helical singularities. Some energy can be saved by deforming the cores of the chiral $Q=-1$ singularities instead of introducing well-separated new cores for helical $\alpha$ singularities.

The main significance of short-scale vortex clusters comes from their classical metastability. Recall that low-energy chiral vortices are made by the winding of either $\theta_\uparrow$ or $\theta_\downarrow$, but not both. Therefore, a chiral vortex has a core in only one component of the order-parameter, corresponding to the angle $\theta_\uparrow$ or $\theta_\downarrow$ that winds. Such a core can be arranged by keeping the overall order-parameter amplitude $\zeta$ constant in (\ref{OP2}), while pinning $\alpha \to (2n+1)\pi$ at the $\theta_\uparrow$ singularities and $\alpha \to 2n\pi$ at the $\theta_\downarrow$ singularities. Apart from these ``boundary conditions'', $\alpha$ varies linearly with the distance from a singularity in its immediate vicinity. An important consequence is that $\alpha$ must change by an integer multiple of $\pi$ on any path between two singularities. If the two singularities are formed in the same angle $\theta_\uparrow$ or $\theta_\downarrow$, then $\alpha$ must change by an integer multiple of $2\pi$ on any path between them. Given that $|\boldsymbol{\nabla}\alpha| \approx \frac{1}{2}mv$ minimizes the Rashba energy, the preferred distance between neighboring singularities is $2\pi(mv)^{-1}n$ where $n$ is an integer. This provides classical metastability to quadruplets, and protects them from immediate annihilation or transformation to large-scale vortex structures. In an attempt to gradually compress a quadruplet and eventually annihilate vortices, one must gradually pay energy for squeezing the period of $\alpha$ oscillations past its optimal value. Energy is ultimately gained only by annihilation, but this is not a classical process.

Given the freedom to make chiral vortices by winding either $\theta_\uparrow$ or $\theta_\downarrow$, it is possible to construct heterogeneous quadruplets that combine singularities in both angles. The sources and drains of the helical currents can reside on the vortices of $\theta_\uparrow$ and $\theta_\downarrow$ respectively. However, the boundary conditions at vortex cores restrict the geometric structure of clusters depending on the types of singularities. For example, the cluster in Fig.\ref{VorQuad} can be formed only from four U(1) singularities of the same angle $\theta_\uparrow$ or $\theta_\downarrow$. This follows from the cluster's symmetry: since $\alpha$ changes by $\pi$ between a vortex and an antivortex, it changes by $2\pi$ between the two vortices. The core boundary condition then implies that the two chiral vortices must be made by winding of the same chiral angle. The two chiral antivortices must unwind the same angle to achieve neutrality. Among all homogeneous quadruplets of the same winding angle, the one in Fig.\ref{VorQuad} has the lowest energy because it maximizes the distance between repelling vortices of the same charge. Heterogeneous clusters shown in Fig.\ref{VorQuad2} have different structures.

\begin{figure}
\includegraphics[width=3.3in]{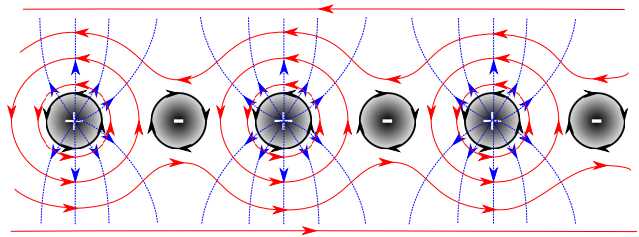}
\caption{\label{VorWall}A domain wall configuration of vortices separating two regions with different uniform type-I order parameters.}
\end{figure}

There are special configurations of closely-packed vortices that cost a finite amount of energy per vortex. They feature the absence of a connected mean-field type-I condensate in their background. The simplest such configuration is a domain wall shown in Fig.\ref{VorWall}. It separates two regions in space with different type-I condensates. Quantum fluctuations necessarily delocalize vortices along the wall, and also prohibit its long-range directional order. Interestingly, however, a domain wall carries a finite amount of secondary charge per unit length, so it cannot self-annihilate even one quadruplet at a time. It takes an ``anti-wall'' (with only $\boldsymbol{\nabla}\alpha$ drains) to annihilate a domain wall (which only has $\boldsymbol{\nabla}\alpha$ sources). Domain walls of delocalized vortices can naturally position themselves at the system boundaries of type-I condensates.

\begin{figure}
\includegraphics[height=2.0in]{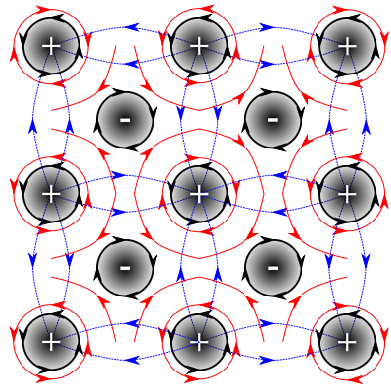}
\caption{\label{VorLat}A TR-invariant lattice of type-I vortices.}
\end{figure}

The most interesting configuration of type-I vortices is a vortex lattice shown in Fig.\ref{VorLat}. It is obtained by tiling the elementary vortex quadruplets as unit-cells. Given that each quadruplet is a metastable structure, the entire vortex lattice is a metastable state itself. Therefore, even when the ``uniform'' (or ``striped'') type-I condensate is the true ground state, this vortex lattice could survive in appropriate conditions for a very long time once it is created. Switching between the ``uniform'' and vortex lattice states would have all properties of a first order phase transition, because these two states correspond to well-defined local minimums separated by a barrier in the energy landscape. First order transitions are not universal, so there is no fundamental obstacle to the existence of microscopic systems in which this vortex lattice is the true ground state. Possible evidence for this is found in recent numerical studies involving lattice models of Rashba spin-orbit-coupled bosons and Cooper-paired fermions \cite{Cole2012, Nikolic2014a}.

\subsection{Type-II vortices}\label{secDoubletVorticesT2}

This section scrutinizes vortex excitations in $S=\frac{1}{2}$ type-II condensates. Qualitatively, all kinds of vortices in type-I states have their analogues in type-II condensates. However, only the pure charge vortices interact through a logarithmic Coulomb potential at large distances, while both the helical and chiral vortices interact linearly and exhibit asymptotic freedom by the string attachment mechanism. There are also notable changes in the vortex dynamics at short distances between vortices: type-II condensates do not support metastable vortex clusters or vortex lattices.

\subsubsection{Large-scale vortex structure}

A type-II state features charge currents and the optimum constant value of $\alpha=\frac{\pi}{2}$. Its energy gain (\ref{Gr1}) through the Rashba spin-orbit coupling is most transparently seen in the (\ref{OP2}) representation of the order parameter:
\begin{eqnarray}\label{Gr3}
E_{\textrm{kin}} \!\!&\propto&\!\! \int \dd^2 r \; \biggl\lbrack (\boldsymbol{\nabla}\zeta)^{2}
  +\zeta^{2}(\boldsymbol{\nabla}\gamma)^{2}+\frac{\zeta^{2}}{4}(\boldsymbol{\nabla}\theta)^{2}
  +\frac{\zeta^{2}}{4}(\boldsymbol{\nabla}\alpha)^{2} \nonumber \\
&& -\zeta^{2}\cos\alpha(\boldsymbol{\nabla}\theta)(\boldsymbol{\nabla}\gamma)
     +\frac{mv\zeta^{2}}{4}(\hat{{\bf z}}\times\hat{\boldsymbol{\theta}})\boldsymbol{\nabla}\alpha \nonumber \\
&& -\frac{mv\zeta^{2}}{2}\sin\alpha\,\hat{\boldsymbol{\theta}}\boldsymbol{\nabla}\gamma \biggr\rbrack \\
&& \!\!\!\!\!\!\!\!\!\!\!\!\!\!\!\!\!\!\!\! \xrightarrow{\alpha\to\frac{\pi}{2}} \int \dd^2 r \; \biggl\lbrack
  (\boldsymbol{\nabla}\zeta)^{2}+\frac{\zeta^{2}}{4}(\boldsymbol{\nabla}\theta)^{2}
  +\zeta^{2}\left(\boldsymbol{\nabla}\gamma-\frac{mv}{4}\hat{\boldsymbol{\theta}}\right)^2 \biggr\rbrack \nonumber
\end{eqnarray}
It is optimal to keep:
\begin{equation}\label{HelicalT2}
\boldsymbol{\nabla}\gamma = \frac{mv}{4}\hat{\boldsymbol{\theta}}
\end{equation}
and thus produce a helical spin current in the background of static spin polarization according to (\ref{CurDen1}), (\ref{CurDen2}) and (\ref{ThetaHat0}).

A pure charge vortex of charge $q\in\mathbb{Z}$ has the configuration $\gamma({\bf r})={\bf kr} + q\phi$ as a function of the polar angle $\phi$, while $\theta=\textrm{const}$. Since $\hat{\boldsymbol{\theta}}$ remains uniform, we can set ${\bf k} = \frac{mv}{4}\hat{\boldsymbol{\theta}}$ and pay only a logarithmically divergent energy $\propto \log(R)$ with the system size $R$ for an isolated vortex. Conversely, two charge vortices interact through a potential that depends logarithmically on the distance between them. The charge vortices of a type-II condensate are equivalent to the helical vortices of type-I condensates.

A helical vortex involves the winding of $\alpha$ by an integer multiple of $2\pi$ on loops that enclose the singularity. Any gradual winding of $\alpha$ that minimizes the energy of $(\boldsymbol{\nabla}\alpha)^2$ unavoidably violates the optimum condition $\alpha=\frac{\pi}{2}$ in the bulk of a type-II state. The ensuing energy cost scales as the system area $R^2$ because the essential Rashba energy gain through the $\sin\alpha\,\hat{\boldsymbol{\theta}}\boldsymbol{\nabla}\gamma$ term of (\ref{Gr3}) is lost on average. No deformation of the uniform charge current and spin polarization can remove this cost without also deforming the configuration of $\alpha$. The only solution is to completely eliminate the bulk $\alpha$ windings and compress them into strings that terminate at the helical singularities. That way, at least, the cost of a single vortex is lowered to something that scales as the system length $R$ instead of area $R^2$. Two helical vortices are connected by a string and interact through a linear potential with the distance between them. In that sense they behave similarly to the chiral vortices of type-I condensates.

The chiral vortices involve circulating ${\bf J}^z \propto \boldsymbol{\nabla}\theta$ spin currents. However, the variable $\theta$ is not adequate for describing quantized chiral vortices, because they must be bound to charge vortices as discussed earlier. We must go back to the representation (\ref{OP2b}) of the order parameter and consider quantized windings in the separate phases $\theta_\uparrow$ and $\theta_\downarrow$ of the two order parameter's components. If we rigidly fix $\alpha=\frac{\pi}{2}$ and $\zeta=\textrm{const}$ far away from any vortex cores, then the energy (\ref{Gr1}) expressed in this representation is:
\begin{equation}\label{Gr4}
E_{\textrm{kin}} \propto \int \dd^2 r \; \frac{\zeta^{2}}{2} \biggl\lbrack
    \left( \boldsymbol{\nabla}\theta_\uparrow - \frac{mv}{4} \hat{\boldsymbol{\theta}} \right)^2
  + \left( \boldsymbol{\nabla}\theta_\downarrow - \frac{mv}{4} \hat{\boldsymbol{\theta}} \right)^2 \biggr\rbrack
\end{equation}
with $\hat{\boldsymbol{\theta}}$ given by (\ref{ThetaHat}). The uniform background charge current ${\bf j} \propto \boldsymbol{\nabla} \gamma = \frac{1}{2} (\boldsymbol{\nabla} \theta_\uparrow + \boldsymbol{\nabla} \theta_\downarrow)$ is established by the concurrent oscillations of $\theta_\uparrow$ and $\theta_\downarrow$ that keep $\theta_\downarrow - \theta_\uparrow$ and thus $\hat{\boldsymbol{\theta}}$ fixed.

An elementary chiral vortex involves $2\pi$ winding in either one of the $\theta_\uparrow, \theta_\downarrow$ angles on a loop that encloses the singularity. This inevitably causes rotations of the vector $\hat{\boldsymbol{\theta}}$ on loops, and thus gives rise to a vortex configuration of the vector field $\hat{\boldsymbol{\theta}}$. We have the same situation as in the type-I condensates: only the $Q=+1$ vector-vortex of $\hat{\boldsymbol{\theta}}$ can be compensated by the gradient of a scalar such as $\boldsymbol{\nabla}\theta_\uparrow$ or $\boldsymbol{\nabla}\theta_\downarrow$ in (\ref{Gr4}). The excessive energy cost of $Q\neq 1$ vortices, which are required for neutrality, can be tamed only by eliminating the winding of $\theta_\uparrow, \theta_\downarrow$ form the bulk and compressing it to strings that emanate from the singularities. The chiral vortices of type-II condensates have the same large-scale dynamics as their type-I analogues.

\subsubsection{Short-scale vortex structure}

The strings between helical or chiral vortices are formed only at length-scales large in comparison to $(mv)^{-1}$. The distribution of spin currents can be more diffused between vortices at distances shorter than $(mv)^{-1}$ from one another. We have seen that this allowed metastable vortex cluster to form in type-I condensates, by the virtue of quantized $\alpha$ windings on the paths \emph{between} singularities. No such mechanism is available in type-II condensates.

Looking at the energy (\ref{Gr3}) of the optimal $\alpha=\frac{\pi}{2}$ type-II condensate, we could attempt to mimic the construction of the type-I metastable clusters. The analogous expression for the energy of type-I states is given by (\ref{Gr2}). It is formally apparent that the charge currents ${\bf j} \propto \boldsymbol{\nabla}\gamma$ of type-II states play the same role as the helical spin currents ${\bf J}^{x,y} \propto \boldsymbol{\nabla}\alpha$ of type-I states. Given a particular configuration of $\hat{\boldsymbol{\theta}}$, the vector field lines of $\boldsymbol{\nabla{\alpha}}$ in the low-energy type-I vortex state should be globally rotated by $\frac{\pi}{2}$ to naively produce the corresponding configuration of $\boldsymbol{\nabla}\gamma$ that costs the same energy in the type-II state. This transforms the type-I source and drain arrangements of helical spin currents into type-II vortex and antivortex patterns of charge currents. This is good because charge currents cannot have sources and drains in equilibrium. We could apply this transformation to the entire configurations shown in Fig.\ref{VorQuad} and \ref{VorQuad2}.

Of course, this construction must be handled carefully to incorporate the quantized windings of $\theta_\uparrow, \theta_\downarrow$. For example, in the type-II analogue of the cluster from Fig.\ref{VorQuad}, one chiral $Q=+1$ vortex should be made by the winding of $\theta_\uparrow$ and the other by the winding of $\theta_\downarrow$. The two chiral $Q=-1$ vortices must also wind different phases. Only this binds chiral and charge vortices in the manner consistent with our mapping. Recall, in contrast, that all four chiral vortices in this cluster must be formed by winding the same phase in type-I condensates.

The short-scale clusters lack metastable rigidity in type-II condensates. The quantized windings of all angles always appear on closed loops that enclose singularities, and never on the open paths between two singularities. Therefore, there is no direct protection of a finite distance between vortices. Bringing the vortices of a neutral cluster closer together always gradually lowers their potential energy until they are annihilated. All small clusters of type-II vortices are short-lived.

The analogy between the type-I and type-II clusters is spoiled in yet another detail. The optimal condition (\ref{HelicalT2}) applied to the circular flows of charge currents necessitates a continuous spatial distribution of the $\textrm{curl}({\bf j}) \neq 0$ vorticity. Since charge vorticity must be quantized in a superfluid state, we cannot perfectly satisfy (\ref{HelicalT2}) even in the vicinity of a $Q=+1$ chiral vortex. There is no such problem with the $Q=+1$ chiral vortices of type-I states. Any chiral singularity of a type-II state must be neutralized at very short length scales in order to avoid the formation of strings.

\section{$S=1$ condensates}\label{secTriplet}

This section is devoted to the two-dimensional condensates of spin triplet particles with a strong Rashba spin-orbit coupling, and their vortex excitations. Triplet condensates can be created directly with bosonic ultra-cold atoms, or indirectly as a result of fermion pairing. The second route can be realized, at least as a matter of principle, in solid-state heterostructure devices that utilize a topological insulator (TI) quantum well. Correlations among the quantum well electrons, which produce triplet pairing in Cooper or exciton channels, must be either artificially engineered by proximity effects, or naturally present in the TI material as in the case of the promising samarium hexaboride (SmB$_6$). 

Following the steps carried out for the two-component bosons in section \ref{secDoublet}, we will first discuss the generic phase diagram of triplet condensates with uniform currents in section \ref{secTripletLG} and identify the same two types of states that gain energy from the Rashba spin-orbit coupling. Then, we will analyze vortex excitations in these two states (sections \ref{secTripletVorticesT1} and \ref{secTripletVorticesT1}). Even though the methodology and qualitative conclusions will be the same as in the case of two-component systems, some details are different and require a separate discussion. Most notably, interesting vortex excitations and lattices can occur without a violation of the TR symmetry.

\subsection{Plain $S=1$ condensates}\label{secTripletLG}

The generic Landau-Ginzburg action of triplet bosons in two spatial dimensions, which respects the TR symmetry, can be written as:
\begin{eqnarray}\label{LG3}
S_{\textrm{t}} \!\! &=& \!\! \int \dd^2 r \, \Biggl\lbrack
      \frac{1}{2m} \Bigl\lbrack\left(\boldsymbol{\nabla}-i\boldsymbol{\mathcal{A}}\right)\eta\Bigr\rbrack^{\dagger}
         \Bigl\lbrack\left(\boldsymbol{\nabla}-i\boldsymbol{\mathcal{A}}\right)\eta\Bigr\rbrack \nonumber \\
  && +t_{\textrm{t}}\eta^{\dagger}\eta +a\eta^{\dagger}\Phi_{0}^{2}\eta \\
  && +U_{\textrm{t}}(\eta^{\dagger}\eta)^{2} +b_1(\eta^{\dagger}\Phi_{0}\eta)^{2} +b_{2}(\eta^{\dagger}\Phi_{0}^{2}\eta)^{2} \Biggr\rbrack
  \ . \nonumber
\end{eqnarray}
The ``external'' SU(2) gauge field $\boldsymbol{\mathcal{A}}$ that embodies the Rashba spin-orbit coupling, and its Yang-Mills flux $\Phi_0$, have the same representation-independent form as (\ref{GaugeField}) and (\ref{Flux}) respectively. Specific to the spin $S=1$ representation are only the spin projection operators that appear in the definition of $\boldsymbol{\mathcal{A}}$:
\begin{eqnarray}
&&
  S^x = \frac{1}{\sqrt{2}}\left(\begin{array}{ccc}0&1&0\\1&0&1\\0&1&0\end{array}\right) \quad,\quad
  S^y = \frac{1}{\sqrt{2}}\left(\begin{array}{ccc}0&-i&0\\i&0&-i\\0&i&0\end{array}\right) \nonumber \\
&& ~~~~~~~~~~~~~~~~~~~
  S^z = \left(\begin{array}{ccc}1&0&0\\0&0&0\\0&0&-1\end{array}\right) \ .
\end{eqnarray}
The action looks more complicated than its doublet counterpart (\ref{LG2}) mainly because $\Phi_0^2 \neq 1$ is not trivial in the $S=1$ representation. If the bosonic degrees of freedom originate from fermion pairing, then strictly speaking the action should contain singlet fields as well, and possibly also Grassmann fields for low energy fermionic excitations. Instead of dealing with such complications, we will focus on situations promoted by the strong spin-orbit coupling in which the dominant low-energy degrees of freedom are spin triplets.

A generic triplet order parameter can be written as:
\begin{equation}\label{OP3}
\eta = \left(\begin{array}{c} \eta_{\uparrow} \\ \eta_{0} \\ \eta_{\downarrow} \end{array}\right)
     = \sqrt{2}\zeta \left(\begin{array}{c} \cos\alpha\,\sin\beta\, e^{-i\theta} \\ i\,\sin\alpha\, e^{-i\omega} \\ \cos\alpha\,\cos\beta\, e^{i\theta}
       \end{array}\right) e^{i\gamma}
\end{equation}
in terms of one amplitude $\zeta$ and five angles $\alpha, \beta, \gamma, \theta, \omega$ that may vary in space and time. The states with uniform currents can be parametrized by nine constants $(\zeta, \beta, \alpha_{0}, \theta_{0}, \omega_{0}, k_{\alpha}, k_{\theta}, k_{\gamma}, k_{\omega})$ according to:
\begin{eqnarray}
\alpha = \alpha_{0}+{\bf k}_{\alpha}{\bf r} &,& \theta = \theta_{0}+{\bf k}_{\theta}{\bf r} \nonumber \\
\omega = \omega_{0}+{\bf k}_{\omega}{\bf r} &,& \gamma = {\bf k}_{\gamma}{\bf r} \ .
\end{eqnarray}
We do not consider oscillatory spatial variations of $\beta$ because the potential and interaction couplings in the action always pin $\beta$ to $\frac{n\pi}{4}$, meaning $|\eta_\uparrow|=|\eta_\downarrow|$ or $\eta_\uparrow\eta_\downarrow=0$. Some of these constants will be redundant or easy to handle, but we anticipate that wavenumbers $k_{\alpha}, k_{\theta}, k_{\gamma}, k_{\omega}$ corresponding to various currents may be non-zero in certain states shaped by the Rashba spin-orbit coupling.

\begin{figure}
\includegraphics[width=2.7in]{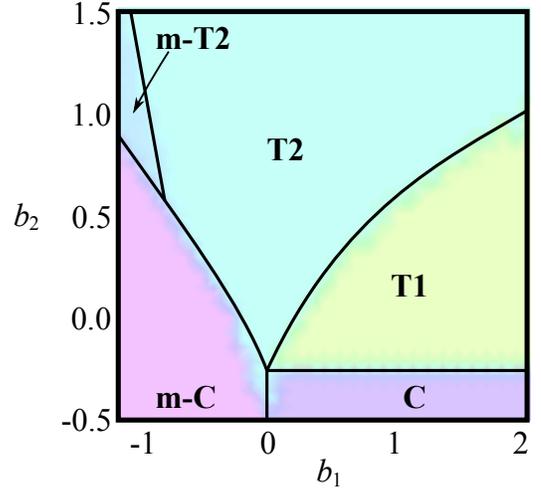}
\caption{\label{LG3pd}The phase diagram of uniform currents in triplet Rashba spin-orbit-coupled condensates. The characteristic types of phases are type-I (T1) and type-2 (T2) condensates that carry spin currents, as well as conventional condensates (C) that carry no currents. A prefix ``m'' indicates finite magnetization in the $z$-direction. This plot was obtained by numerical minimization of the energy functional (\ref{LG3}) with respect to the nine parameters in (\ref{OP3}), at $m=1$, $v=1$, $t_{\textrm{t}}=-2.5$, $a=-0.3$, $U_{\textrm{t}}=2$ in arbitrary units. A qualitatively similar layout of condensates is obtained for all other combinations of coupling constants. As $a$ is increased, the conventional condensates are pushed to smaller values of $b_2$ by the expanding T1 and T2 phases.}
\end{figure}

The phase diagram of condensates with uniform densities or currents is shown in Fig.\ref{LG3pd}. It features two prominent phases that take advantage of the Rashba spin-orbit coupling. They are characterized by the same type-I or type-II patterns of helical spin currents that we discovered in the two-component condensates. The type-I state is a pure flow of helical spin with ${\bf k}_\alpha \neq 0$ and ${\bf k}_\gamma =  {\bf k}_\theta = {\bf k}_\omega = 0$. It respects the TR-symmetry in the $S=1$ systems, unless a spontaneous spin density-wave develops according to the discussion in appendix \ref{secDensityMod}. The type-II state is the flow of charge in the background of uniform spin polarization, which gives rise to the helical spin current at the expense of breaking the TR symmetry. In this state, ${\bf k}_\gamma \neq 0$ and ${\bf k}_\alpha =  {\bf k}_\theta = {\bf k}_\omega = 0$. Either kind of condensates can in principle coexist with additional spin polarization along the $z$-direction, but no type-I states with such a magnetization are found. Similarly, mixtures of type-I and type-II condensates seem to never minimize energy. In some special situations, a conventional condensate without any current flow is stable, but it gains nothing from the Rashba term.

Some qualitative features of the phase diagram can be understood analytically. The potential energy density of triplets expressed in terms of the individual spinor components is:
\begin{eqnarray}\label{Etri}
\mathcal{E}_{\textrm{t}} \!\!&=&\!\! t_{\textrm{t}0}|\eta_{0}|^{2}+t_{\textrm{t}1}\left(|\eta_\uparrow|^{2}+|\eta_\downarrow|^{2}\right) \\
\!\!&+&\!\! U_{\textrm{t}}|\eta_{0}|^{4}
       +\left(U_{\textrm{t}}+b_{1}\Phi^{2}+b_{2}\Phi^{4}\right)\left(|\eta_\uparrow|^{2}+|\eta_\downarrow|^{2}\right)^{2} \nonumber \\
\!\!&-&\!\! 4b_{1}\Phi^{2}|\eta_\uparrow|^{2}|\eta_\downarrow|^{2} \ , \nonumber
\end{eqnarray}
where we defined the effective quadratic couplings for spinless and spinful triplets
\begin{eqnarray}
t_{\textrm{t}0} \!\!&=&\!\! t_{\textrm{t}} + mv^2 \\
t_{\textrm{t}1} \!\!&=&\!\! t_{\textrm{t}} + \frac{1}{2}mv^2 + a\Phi^{2}+2U_{\textrm{t}}|\eta_{0}|^{2} \ . \nonumber
\end{eqnarray}
Spinless $\eta_0$ and spinful $\eta_\uparrow, \eta_\downarrow$ triplets will generally condense when $t_{\textrm{t}0}$ and $t_{\textrm{t}1}$ are negative respectively. The stability condition for the condensate is then:
\begin{equation}
U_{\textrm{t}}>0\quad,\quad U_{\textrm{t}}+b_{2}\Phi^{4}>0\quad,\quad U_{\textrm{t}}+b_{1}\Phi^{2}+b_{2}\Phi^{4}>0 \ . \nonumber
\end{equation}
This allows the coupling $b_1$ to be negative, in which case the term  $-4b_{1}\Phi^{2}\left(|\eta_\uparrow|^{2}|\eta_\downarrow|^{2}\right)$ prefers that one of 
$\eta_\uparrow,\eta_\downarrow$ be zero for any given value of $|\eta_\uparrow|^{2}+|\eta_\downarrow|^{2}$. The ensuing condensates with spontaneous magnetization along the $z$-axis are indeed found numerically only for $b_1<0$. It can be also easily seen that larger values of $a$ and $b_2$ promote the $\eta_0$ spinor component at the expense of $\eta_\uparrow, \eta_\downarrow$. This generally discourages the type-I condensates because their TR-invariant helical spin currents require oscillatory spatial variations of $\alpha$ in (\ref{OP3}), and thus $|\eta_\uparrow| = |\eta_\downarrow| \sim |\eta_0|$ on average. In such circumstances, the type-II condensate has an advantage at any $v>0$. Their charge current (proportional to $\boldsymbol{\nabla}\gamma$) and spin polarization (proportional to $\sin(2\alpha))$ can gain Rashba energy even when only the $\eta_0$ amplitude is large.

A condensate respects the TR symmetry if $\eta_\uparrow^{\phantom{*}} = \eta_\downarrow^*$ and $\eta_0^{\phantom{*}} = -\eta_0^*$ at every point in space. Multiplying such a triplet spinor by a constant global phase factor $e^{i\gamma_0}$ has no measurable consequences and yields the most general TR-invariant order parameter. The condition $|\eta_\uparrow|=|\eta_\downarrow|$ favored by $b_1>0$ fixes $\beta$ to $\frac{\pi}{4}$ in (\ref{OP3}), and then it takes only $\omega=0$ and $\gamma=\textrm{const}$ to have a TR-invariant state. Spatial variations of $\omega$ turn out not to be an efficient way to gain energy from the spin-orbit coupling, so the system prefers to save kinetic energy by $\omega=\textrm{const}$. Instead, the Rashba energy is lowered best by helical spin currents of the type-I or type-II. The former is very naturally consistent with the conditions created by $b_1>0$, and this is where such phases are found in the phase diagram. Type-II states involve spatial variations of $\gamma$, which do not contradict $|\eta_\uparrow|=|\eta_\downarrow|$ required by $b_1>0$, and minimize both the kinetic and Rashba energy by $\omega=\frac{\pi}{2}$.

\subsection{Type-I vortices}\label{secTripletVorticesT1}

Here we explore the helical, chiral and charge vortices in the $S=1$ systems by comparing them to the equivalent excitations in the $S=\frac{1}{2}$ condensates. Helical, chiral and charge vortices are U(1) singularities in the phases $\alpha$, $\theta$ and $\gamma$ respectively. They involve circular currents of different kinds, among which only the charge currents of $\gamma$ singularities violate the TR symmetry. Triplet systems allow additional singular structures of currents, but their dynamics, discussed in section \ref{secExtraSing}, is suppressed by higher energy costs.

The order parameter of an ideal triplet type-I condensate is TR-invariant, and hence can be generally written as:
\begin{equation}\label{OP3T1}
\eta = \left(\begin{array}{c} \eta_{\uparrow} \\ \eta_{0} \\ \eta_{\downarrow} \end{array}\right)
     = \zeta \left(\begin{array}{c} \cos\alpha\, e^{-i\theta} \\ i\sqrt{2}\,\sin\alpha \\ \cos\alpha\, e^{i\theta} \end{array}\right) \ .
\end{equation}
The non-vanishing current density components in this condensate are only those that remain invariant under TR:
\begin{eqnarray}\label{CurrentsT1}
j_0 \!\!&=&\!\! 2 \zeta^2 \\[0.1in]
{\bf J}^{x} \!\!&=&\!\! \frac{2\zeta^{2}}{m}\left\lbrack \cos\theta\,\boldsymbol{\nabla}\alpha
  +\frac{1}{2}\sin\theta\sin(2\alpha)\,\boldsymbol{\nabla}\theta\right\rbrack \nonumber \\
{\bf J}^{y} \!\!&=&\!\! \frac{2\zeta^{2}}{m}\left\lbrack \sin\theta\,\boldsymbol{\nabla}\alpha
  -\frac{1}{2}\cos\theta\sin(2\alpha)\,\boldsymbol{\nabla}\theta\right\rbrack \nonumber \\
{\bf J}^{z} \!\!&=&\!\! -\frac{2\zeta^{2}}{m}\cos^{2}\alpha\,\boldsymbol{\nabla}\theta \nonumber \ .
\end{eqnarray}
Specifically, there is no spin-texture ($J_0^a = 0$) and no flow of charge (${\bf j}=0$), but charge density $j_0 \neq 0$ is free to break translational symmetry, and so is the spin-current density ${\bf J}^a$. The gradient energy of the order parameter (\ref{OP3T1}) extracted from (\ref{LG3}) is:
\begin{eqnarray}\label{Kin2}
\!\!\!\!\!\!\!\! E_{\textrm{kin}} \!\!&=&\!\! \int \dd^2 r \; \Bigl\lbrace \frac{1}{2m} |\boldsymbol{\nabla} \eta|^2
   + mv \left( J_y^x - J_x^y \right) \Bigr\rbrace \\
\!\!&\propto&\!\! \int \dd^2 r \Bigl\lbrace (\boldsymbol{\nabla}\zeta)^{2}+\zeta^{2}(\boldsymbol{\nabla}\alpha)^{2}
      +\zeta^{2}\cos^{2}\alpha\,(\boldsymbol{\nabla}\theta)^{2} \nonumber \\
&&  +mv\zeta^{2}\Bigl\lbrack\sin(2\alpha)\boldsymbol{\nabla}\theta
      -2(\hat{{\bf z}}\times\boldsymbol{\nabla}\alpha)\Bigr\rbrack\hat{\boldsymbol{\theta}}\Bigr\rbrace \nonumber \ ,
\end{eqnarray}
where the unit-vector
\begin{equation}\label{ThetaHat1}
\hat{\boldsymbol{\theta}} = \hat{\bf x}\cos\theta + \hat{\bf y}\sin\theta 
\end{equation}
represents the direction of spin that is transported by a helical current ${\bf J}^{x,y} \sim {\boldsymbol{\nabla}\alpha}$. 

A type-I state features rapid spatial oscillations of the angle $\alpha$, with the period $\sim(mv)^{-1}$. The TR-invariant dynamics at length-scales much larger than $(mv)^{-1}$ is qualitatively captured by the much simpler coarse-grained energy:
\begin{equation}\label{Kin2b}
E_{\textrm{kin}} =\! \int \dd^2 r \left\lbrack (\boldsymbol{\nabla}\zeta)^{2}
    +\zeta^{2}(\boldsymbol{\nabla}\alpha+mv\,\hat{{\bf z}}\times\hat{\boldsymbol{\theta}})^{2}
      +\frac{\zeta^{2}}{2}(\boldsymbol{\nabla}\theta)^{2} \right\rbrack
\end{equation}
The analogy to the dynamics of the $S=\frac{1}{2}$ type-I condensates is immediately apparent if we compare this expression with  (\ref{Gr2}). There are only two notable differences between the $S=1$ and $S=\frac{1}{2}$ cases. The optimal magnitude $|\boldsymbol{\nabla}\alpha| = mv$ has changed, merely due to the larger spin representation. More importantly, the chiral phase $\theta$ of the $S=1$ systems can freely wind by $2\pi$ without introducing line discontinuities in the order parameter. This frees the chiral vortices from the binding to charge vortices, and allows them to respect the TR symmetry. There is only one flavor of chiral vortices, as opposed to two in the $S=\frac{1}{2}$ systems. \footnote{The two flavors of chiral vortices in the $S=\frac{1}{2}$ systems are associated to the winding of $\theta_\uparrow$ or $\theta_\downarrow$. The formally equivalent flavors in the $S=1$ systems would involve the variations of $\beta$, which are strongly suppressed (see section \ref{secExtraSing}).}

Apart from these differences, it is clear by analogy that the helical and chiral vortices of the type-I $S=1$ systems have the same large-scale dynamics as their $S=\frac{1}{2}$ counterparts. The helical vortices cost energy that scales as $\log(R)$ with the system size, while the energy of chiral vortices scales as $R$. The potential energy of interaction between two vortices separated by $l$ is proportional to $\log(l)$ in the case of helical and $l$ in the case of chiral vortices. The latter are connected by string-like regions of compressed order parameter modulations whose thickness is of the order of $(mv)^{-1}$. The expressions (\ref{Kin2}) and (\ref{Kin2b}) are specialized for the TR-invariant order parameters, so they hide the energy of charge vortices. Nevertheless, charge currents gain nothing from the Rashba spin-orbit coupling on average in the type-I states, so the cost of charge vortices is logarithmic in the system size.

Since the optimum condition $\boldsymbol{\nabla}\alpha = -mv\,\hat{{\bf z}}\times\hat{\boldsymbol{\theta}}$ for the low-energy helical currents emerges directly from (\ref{Kin2}), it holds microscopically and independently of coarse-graining. We showed that the equivalent condition enables small metastable clusters of chiral vortices in the $S=\frac{1}{2}$ systems, so by analogy such clusters exist in the $S=1$ systems as well. Metastability is now established by the fact that the winding of $\theta$ wants to deplete the vortex cores in the $\eta_\uparrow$ and $\eta_\downarrow$ components of the order parameter (\ref{OP3}), while it leaves alone the $\eta_0$ component. Energy is saved by keeping the amplitude $\zeta$ (and thus $|\eta_0|$) finite inside the core, and pining $\alpha$ to $\frac{\pi}{2} + n\pi$ at the singularity to carve cores only in the $\eta_\uparrow$ and $\eta_\downarrow$ components. Therefore, $\alpha$ must change by an integer multiple of $\pi$ on any path between two singularities, which in turn tends to keep an optimum distance $\sim n(mv)^{-1}$ between them according to $|\boldsymbol{\nabla}\alpha|\sim mv$.

Again, specific to the $S=1$ systems is a single flavor of chiral vortices. There is only one metastable chiral quadruplet whose pattern of currents is shown in Fig.\ref{VorQuad} or \ref{VorQuad2}. Similarly, domain walls shown in Fig.\ref{VorWall} and vortex lattices shown in Fig.\ref{VorLat} are unique.

\subsection{Type-II vortices}\label{secTripletVorticesT2}

Here we show that the properties of helical, chiral and charge vortices in $S=1$ type-II condensates are analogous to those of the two-component systems.

The general type-II order parameter without magnetization along the $z$-axis can be written as:
\begin{equation}\label{OP3T2}
\eta = \left(\begin{array}{c} \eta_{\uparrow} \\ \eta_{0} \\ \eta_{\downarrow} \end{array}\right)
  = \zeta\left(\begin{array}{c} \cos\alpha\,e^{-i\theta} \\ \sqrt{2}\,\sin\alpha \\ \cos\alpha\,e^{i\theta} \end{array}\right) e^{i\gamma}
  \ .
\end{equation}
This time we will treat $\gamma$ and $\theta$ as variable U(1) phases, and assume at first that $\alpha$ is a constant whose optimal value we need to determine. The charge density and currents are then:
\begin{eqnarray}
j_{0} &=& 2\zeta^{2} \\
J_{0}^{x} &=& 2\zeta^{2}\sin(2\alpha)\cos\theta \nonumber \\
J_{0}^{y} &=& 2\zeta^{2}\sin(2\alpha)\sin\theta \nonumber \\
J_{0}^{z} &=& 0 \nonumber \ ,
\end{eqnarray}
while the spin density and currents are:
\begin{eqnarray}
{\bf j} &=& \frac{2\zeta^{2}}{m}(\boldsymbol{\nabla}\gamma) \\
{\bf J}^{x} &=& \frac{2\zeta^{2}}{m}\sin(2\alpha)\cos\theta\,(\boldsymbol{\nabla}\gamma) \nonumber \\
{\bf J}^{y} &=& \frac{2\zeta^{2}}{m}\sin(2\alpha)\sin\theta\,(\boldsymbol{\nabla}\gamma) \nonumber \\
{\bf J}^{z} &=& -\frac{2\zeta^{2}}{m}\cos^{2}\!\alpha\,(\boldsymbol{\nabla}\theta) \nonumber \ .
\end{eqnarray}
The gradient energy of the order parameter (\ref{OP3T2}) obtained from (\ref{LG3}) is:
\begin{eqnarray}\label{Kin3}
E_{\textrm{kin}} \!\!&=&\!\! \int \dd^2 r \; \Bigl\lbrace \frac{1}{2m} |\boldsymbol{\nabla} \eta|^2
   + mv \left( J_y^x - J_x^y \right) \Bigr\rbrace \\
\!\!&\propto&\!\! \int \dd^2 r \Bigl\lbrace (\boldsymbol{\nabla}\zeta)^{2}
      +\zeta^{2}\cos^{2}\alpha\,(\boldsymbol{\nabla}\theta)^{2} \nonumber \\
\!\!&+&\!\! \zeta^2 \Bigl( \boldsymbol{\nabla}\gamma + mv\sin(2\alpha)\, \hat{{\bf z}}\times\hat{\boldsymbol{\theta}} \Bigr)^2
      -(mv\zeta)^2 \sin^2(2\alpha) \Bigr\rbrace \ . \nonumber
\end{eqnarray}
It is immediately clear from the last term that the optimum values for $\alpha$ are $\pm\pi/4$, $\pm 3\pi/4$. Then, it pays to keep $\hat{\boldsymbol{\theta}}$ perpendicular to $\boldsymbol{\nabla}\gamma$, which amounts to a background uniform spin polarization perpendicular to the flow of charge current. Now, $\hat{\boldsymbol{\theta}}$ given by (\ref{ThetaHat1}) represents the direction of spin polarization. 

The energy (\ref{Kin3}) depends on $\alpha$, $\theta$ and $\gamma$ in qualitatively the same manner as the energy (\ref{Gr3}) of the analogous $S=\frac{1}{2}$ type-II state, up to a global $\frac{\pi}{2}$ rotation of $\theta$. We can immediately conclude that the helical, chiral and charge vortices exhibit the same kind of dynamics in the $S=1$ and $S=\frac{1}{2}$ type-II states. This is true both at large and short distances in comparison to $(mv)^{-1}$ between vortices. Again, the only notable difference is that the $S=1$ systems support only one flavor of chiral vortices (see previous section for details).

\medskip
\subsection{Additional singularities of currents}\label{secExtraSing}

In addition to helical, chiral and charge singularities of the phases $\alpha$, $\theta$ and $\gamma$ respectively, the order parameter (\ref{OP3}) is expressed in terms of two more angles, $\omega$ and $\beta$, which can have their own U(1) singularities. Here we briefly comment on the properties of these vortex excitations.

We already emphasized in section \ref{secTripletLG} that the Landau-Ginzburg action (\ref{LG3}) promotes ground states in which either $|\eta_\uparrow| = |\eta_\downarrow|$, or at least one of the amplitudes $\eta_\uparrow$, $\eta_\downarrow$ is zero. Both cases pin the value of $\beta$ to a constant. This has nothing to do with the Rashba spin-orbit coupling and applies to all TR-invariant triplet systems. Any spatial variation of $\beta$ away from its optimal value for the established condensate costs a finite energy per unit area. Therefore, the conventional vortices of $\beta$ cost energy that scales as the system area $R^2$. We can fix this problem using the same trick as before. If we keep $\beta$ uniform in the bulk and allow it to wind only across thin strings that connect vortices, then the energy cost is reduced to something proportional to the string length. Still, even such singularities are more costly than the similar chiral ones. To see this, consider the full expression for the kinetic and Rashba energy density of the order parameter (\ref{OP3}):
\begin{widetext}
\begin{eqnarray}\label{KinEnergy}
\mathcal{E}\!\!&=&\!\!
  \frac{\zeta^{2}}{2m}\Bigl\lbrace(mv)^{2}\Bigl\lbrack3-\cos(2\alpha)\Bigr\rbrack+2(\boldsymbol{\nabla}\alpha)^{2}
  +2(\boldsymbol{\nabla}\gamma)^{2}+2\cos^{2}\alpha\Bigl\lbrack(\boldsymbol{\nabla}\beta)^{2}+(\boldsymbol{\nabla}\theta)^{2}\Bigr\rbrack
  -4\sin^{2}\alpha\,(\boldsymbol{\nabla}\gamma)(\boldsymbol{\nabla}\omega)
  +2\sin^{2}\alpha\,(\boldsymbol{\nabla}\omega)^{2}\Bigr\rbrace \nonumber \\
&&	+\frac{v\zeta^{2}}{\sqrt{2}}\cos\beta\Bigl\lbrace2(\hat{{\bf z}}\times\hat{\boldsymbol{\theta}}_{+})
	(\boldsymbol{\nabla}\alpha)-\sin(2\alpha)(\hat{{\bf z}}\times\hat{\boldsymbol{\theta}}_{-})
	(\boldsymbol{\nabla}\beta)+\sin(2\alpha)(2\boldsymbol{\nabla}\gamma
	+\boldsymbol{\nabla}\theta-\boldsymbol{\nabla}\omega)\hat{\boldsymbol{\theta}}_{+}\Bigr\rbrace \nonumber \\
&&	+\frac{v\zeta^{2}}{\sqrt{2}}\sin\beta\Bigl\lbrace2(\hat{{\bf z}}\times\hat{\boldsymbol{\theta}}_{-})
	(\boldsymbol{\nabla}\alpha)+\sin(2\alpha)(\hat{{\bf z}}\times\hat{\boldsymbol{\theta}}_{+})(\boldsymbol{\nabla}\beta)+\sin(2\alpha)
	(-2\boldsymbol{\nabla}\gamma+\boldsymbol{\nabla}\theta+\boldsymbol{\nabla}\omega)\hat{\boldsymbol{\theta}}_{-}\Bigr\rbrace \\
&&	+\frac{2\zeta^{2}}{m}\cos^{2}\alpha\,\cos(2\beta)(\boldsymbol{\nabla}\gamma)(\boldsymbol{\nabla}\theta)+\frac{1}{m}
	(\boldsymbol{\nabla}\zeta)^{2} \ , \nonumber
\end{eqnarray}
\end{widetext}
where
\begin{equation}
\hat{\boldsymbol{\theta}}_{\pm}=\hat{{\bf x}}\cos(\theta\pm\omega)+\hat{{\bf y}}\sin(\theta\pm\omega) \ .
\end{equation}
Clearly, the oscillatory variation of $\beta$ averages out to zero the Rashba energy gain (in the middle two lines) acquired through any type of current. Hence, the $\beta$-vortices cost not only the interaction energy but also the Rashba energy. Other expensive vortices typically cost only the Rashba energy.

Now we turn to the singularities of $\omega$. We may coarse-grain (\ref{KinEnergy}) in the type-I states to see more clearly the effect of $\omega$ variations. All Rashba terms proportional to $\sin(2\alpha)$ average out to zero, but
\begin{equation}
(\hat{{\bf z}}\times\hat{\boldsymbol{\theta}}_{+})(\boldsymbol{\nabla}\alpha)
  + (\hat{{\bf z}}\times\hat{\boldsymbol{\theta}}_{-})(\boldsymbol{\nabla}\alpha)
= 2\cos\omega \, (\hat{{\bf z}}\times\hat{\boldsymbol{\theta}})(\boldsymbol{\nabla}\alpha)
\end{equation}
survives, where $\hat{\boldsymbol{\theta}}$ is given by (\ref{ThetaHat1}). Therefore, the oscillations of $\omega$ completely remove the Rashba energy gain on average in type-I states. Conventional U(1) singularities of $\omega$ cost a finite energy per unit area. This can be improved by attaching strings to $\omega$-vortices, but no short-scale structures like those of chiral vortices can be formed because $\omega$ is completely decoupled from $\hat{\boldsymbol{\theta}}$.

The properties of $\omega$-vortices in symmetric type-II condensates can be revealed by setting $\alpha=\beta=\frac{\pi}{4}$ and keeping $\theta$ fixed. The Rashba energy is proportional to:
\begin{equation}
(2\boldsymbol{\nabla}\gamma-\boldsymbol{\nabla}\omega)(\hat{\boldsymbol{\theta}}_{+}-\hat{\boldsymbol{\theta}}_{-})
 = 2\sin\omega\,(2\boldsymbol{\nabla}\gamma-\boldsymbol{\nabla}\omega) (\hat{{\bf z}}\times\hat{\boldsymbol{\theta}}) \ ,
\end{equation}
so again the oscillations of $\omega$ kill the Rashba energy gain and cost a finite energy density.

At the end, let us recall that the phase diagram of triplet bosons contains type-II condensates with magnetization along the $z$-direction. It turns out that this magnetization is never saturated in the type-II state, i.e. $\beta$ changes gradually away from the extreme values $\frac{n\pi}{2}$. That being the case, all conclusions we reached about the dynamics of vortices in symmetric type-II states (with $\beta=\frac{\pi}{4}$) hold unchanged irrespective of the $z$-magnetization. Saturated magnetization in the $z$-direction is found only in the m-C phase of Fig.\ref{LG3pd}, but that phase takes no advantage of the Rashba spin-orbit coupling.

\section{Conclusions}\label{secConclusions}

We studied two- and three-component SU(2) condensates in the presence of a strong Rashba spin-orbit coupling in two spatial dimensions. Our goal was to classify and characterize the U(1) singularities of circulating currents. For this purpose, we used the plain currents (\ref{Currents}) rather than the gauge-covariant ones (\ref{GCCur}). Table \ref{Sum1} summarizes the mean-field phases, and table \ref{Sum2} summarizes the types and properties of vortex singularities that are common to both two- and three-component systems.

\begin{table*}[!]
\centering
\begin{tabular}{||l||c|c|c|c|c|c||c|c|c|c|c|c||}
\hline\hline
type of system & \multicolumn{6}{c||}{two-component $S=\frac{1}{2}$} & \multicolumn{6}{c||}{three-component $S=1$} \\
\hline
phase & type-I & type-II & M-type-II & conv. & M-conv. & normal & type-I & type-II & M-type-II & conv. & M-conv. & normal \\
\hline\hline
TR-invariant & no & no & - & - & no & yes & yes$^*$ & no & no & yes & no & yes \\
\hline
translation-invariant & no & yes & - & - & yes & yes & yes$^*$ & yes & yes & yes & yes & yes \\
\hline
rotation-invariant & no & no & - & - & yes & yes & no & no & no & yes & yes & yes \\
\hline
superfluid density & finite & finite & - & - & finite & 0  & finite & finite & finite & finite & finite & 0 \\
\hline
spin density $\langle S^z\rangle$ & oscil. & 0 & - & - & finite & 0  & 0 & 0 & finite & 0 & finite & 0\\
\hline
spin density $\langle S^{x,y}\rangle$ & oscil. & finite & - & - & 0 & 0  & 0$^*$ & finite & 0 & 0 & 0 & 0\\
\hline
charge current $\langle{\bf j}\rangle$ & 0 & finite & - & - & 0 & 0  & 0 & finite & finite & 0 & 0 & 0\\
\hline
chiral spin current $\langle {\bf J}^z\rangle$ & 0 & 0 & - & - & 0 & 0  & 0 & 0 & finite & 0 & 0 & 0\\
\hline
helical spin current $\langle {\bf J}^{x,y}\rangle$ & finite & finite & - & - & 0 & 0  & finite & finite & finite & 0 & 0 & 0\\
\hline\hline
\end{tabular}
\caption{\label{Sum1}The summary of the plain mean-field phases and the qualitative properties of their currents as defined by (\ref{Currents}). An asterisk indicates a conditionally realized property of the charge and spin current/density patterns, not necessarily the phase as a whole (type-I condensates in higher spin representations typically have ``hidden'' density-wave orders).}
\end{table*}

\begin{table*}[!]
\centering
\begin{tabular}{||l||c|c|c||c|c|c||}
\hline\hline
condensate & \multicolumn{3}{c||}{type-I} & \multicolumn{3}{c||}{type-II} \\
\hline
vortex & helical & chiral & charge & helical & chiral & charge \\
\hline\hline
type of current & in-plane spin ${\bf J}^{x,y}$ & $z$-spin ${\bf J}^z$ & charge ${\bf j}$ &
  in-plane spin ${\bf J}^{x,y}$ & $z$-spin ${\bf J}^z$ & charge ${\bf j}$ \\
\hline
energy scaling & $\log(R)$ & $R$ & $\log(R)$ & $R$ & $R$ & $\log(R)$ \\
\hline
interactions & Coulomb & strong & Coulomb & strong & strong & Coulomb \\
\hline
correlations & - & \multicolumn{2}{c||}{bound if $S=\frac{1}{2}$} & - & \multicolumn{2}{c||}{bound if $S=\frac{1}{2}$} \\
\hline
metastable structures & \multicolumn{2}{c|}{quadruplets, lattices, etc.} & no & no & no & no \\
\hline\hline
\end{tabular}
\caption{\label{Sum2}The summary of quantized U(1) singularities common to the two- and three-component condensates. Strong interactions indicate confinement and asymptotic freedom, and $R$ is the linear system size. The three-component type-I and type-II condensates feature two additional types of high-energy strongly-interacting U(1) vortices.}
\end{table*}

There are two notable features of these systems not found in conventional superfluids. First, some types of vortices interact via linear potentials as a function of distance. They exhibit confinement just like quarks in quantum chromodynamics. Any attempt to separate a vortex and an antivortex to a large distance results first in a string of high energy density stretched between them. This string will rupture, if it becomes too long, into another vortex-antivortex pair that migrate in the opposite directions to screen-out the original singularities. Second, one of the mean-field phases has unusual metastable structures of vortices and antivortices: quadruplets (Fig.\ref{VorQuad}), domain walls (Fig.\ref{VorWall}) and vortex lattices (Fig.\ref{VorLat}). Quadruplets are formed from confined vortex types within the spatial range of their asymptotic freedom. Metastable vortex lattices of tiled quadruplets can become stable in particular microscopic systems, such as tight-binding crystals (which will be studied elsewhere). Such vortex lattices are especially interesting because their quantum melting can give rise to fractional topological insulator states with novel properties.

\section{Acknowledgements}

This work was supported by the National Science Foundation under Grant No. PHY-1205571, and also in part by the National Science Foundation under Grant No. PHY-1066293 with hospitality of the Aspen Center for Physics.

\appendix

\section{Current conservation laws}\label{ConsLaws}

In this technical appendix we scrutinize the discovered vortex properties from the current conservation point of view. In principle, detailed order parameter configurations in the vicinity of vortices can be calculated by solving the current conservation equations. This task is unfortunately too difficult in general circumstances, so our goal is mainly to show that our conclusions are consistent with equilibrium requirements. For example, the singular type-I patterns have sources of helical spin currents, which turn out to be allowed (and necessary) unlike the sources of charge currents. We will also discuss the intricate spin density modulations of the order parameter at short length-scales comparable to $(mv)^{-1}$.

The superfluid order parameter is a classical quantity by the virtue of being the expectation value of a field operator. Therefore, the currents obtained from the order parameter must obey the classical conservation laws. Spin-current conservation in general theories of particles coupled to SU(2) gauge fields can be expressed as:
\begin{equation}\label{CC}
\partial_\mu I_\mu -i\lbrack\mathcal{A}_\mu,I_\mu\rbrack + i\left(\eta^{\dagger}\lbrack S^{a}, \delta H \rbrack\eta\right)S^{a} = 0
\end{equation}
where $I_\mu^{\phantom{a}} = I_\mu^a S^a$ are the SU(2) matrices of the gauge covariant spin-currents
\begin{equation}\label{GCCur}
I_0^a = J_0^a \quad,\quad I_i^a = J_i^a - \frac{1}{2m}\eta^\dagger\lbrace S^a,\mathcal{A}_i\rbrace\eta \ .
\end{equation}
and $J_\mu^a$ are given by (\ref{Currents}). Spin-current conservation is affected by the non-minimal couplings of matter to the external spin-orbit gauge field, which we collected in the $\delta H$ part of the Hamiltonian. For example, the relevant non-minimal couplings in the $S=1$ representation of SU(2) included in (\ref{LG3}) are:
\begin{equation}\label{NonMin}
\delta H = a\Phi_{0}^{2} + b_{1}\left(\eta^{\dagger}\Phi_{0}^{\phantom{\dagger}}\eta\right)\Phi_{0}^{\phantom{\dagger}}
  + b_{2}\left(\eta^{\dagger}\Phi_{0}^{2}\eta\right)\Phi_{0}^{2} \ .
\end{equation}
They arise in our treatment because the external spin-orbit SU(2) gauge field has a non-zero flux $\Phi_0$, unlike for example the Yang-Mills theories in high-energy physics. Their main effects are to bias the ground-state toward one of the type-I, type-II or more conventional condensates, as seen in section \ref{secTripletLG}, and to shape the short-scale modulations of the order parameter that we discuss in subsection \ref{secDensityMod}. Here, we will merely assume that the resolution of this bias is either a type-I or type-II state, and examine the remaining constraints on its order parameter by setting $\delta H \to 0$ in (\ref{CC}). Our qualitative conclusions in this section are immune to this potentially large approximation. The remaining expression (\ref{CC}) with only the minimal coupling is perhaps more familiar in its gauge-covariant form $\lbrack D_\mu, I_\mu \rbrack = 0$, where $D_\mu = \partial_\mu - i\mathcal{A}_\mu$ is the covariant derivative. Substituting $S^a$ by the identity matrix in the above formulas reveals the charge conservation law:
\begin{equation}\label{ChargeCons}
\partial_\mu i_\mu = 0 \quad \cdots \quad
  i_0 = j_0 \quad,\quad i_i = j_i - \frac{1}{m}\eta^\dagger\mathcal{A}_i\eta \ .
\end{equation}

Current conservation laws can be derived from the equation of motion, which for a non-relativistic theory like ours is the Schrodinger equation (or its adjoint):
\begin{eqnarray}
\frac{1}{2m}(-i\boldsymbol{\nabla}-\boldsymbol{\mathcal{A}})^{2}\eta + \delta H \, \eta \!\!&=&\!\!
  i\frac{\partial\eta}{\partial t} \\
\frac{1}{2m}\Bigl\lbrack(-i\boldsymbol{\nabla}-\boldsymbol{\mathcal{A}})^{2}\eta\Bigr\rbrack^\dagger + \eta^\dagger \delta H \!\!&=&\!\!
 -i\frac{\partial\eta^{\dagger}}{\partial t} \ . \nonumber
\end{eqnarray}
The conservation laws for spin-currents are obtained by taking the difference between the first equation multiplied from left by $\eta^{\dagger}S^{a}$ and the second equation multiplied from right by $S^{a}\eta$. After some algebraic manipulation, one arrives at (\ref{CC}).

All time derivatives in the current conservation laws must vanish in equilibrium. Then, the Rashba spin-orbit coupling (\ref{GaugeField}) turns (\ref{CC}) into:
\begin{eqnarray}\label{CurCons2}
\boldsymbol{\nabla}{\bf I}^{x} \!\!&=&\!\!
   -{\bf A}^{y}{\bf I}^{z}+{\bf A}^{z}{\bf I}^{y}=-mv\, I_{x}^{z}\\
\boldsymbol{\nabla}{\bf I}^{y} \!\!&=&\!\!
   -{\bf A}^{z}{\bf I}^{x}+{\bf A}^{x}{\bf I}^{z}=-mv\, I_{y}^{z}\nonumber\\
\boldsymbol{\nabla}{\bf I}^{z} \!\!&=&\!\!
   -{\bf A}^{x}{\bf I}^{y}+{\bf A}^{y}{\bf I}^{x}=mv(I_{x}^{x}+I_{y}^{y})\nonumber \ .
\end{eqnarray}
The scalar gauge field components $A_\mu^a$ are extracted from $\mathcal{A}_\mu^{\phantom{a}} = A_\mu^a S^a$, and we used the relationship $\lbrack S^a, S^b \rbrack = i\epsilon^{abc} S^c$ between the SU(2) generators $S^a$ in any representation. These equations define constraints that the order parameter must satisfy if it is to be static.

In order to express (\ref{CurCons2}) in a relatively compact form, let us define the in-plane spin-currents as double vectors:
\begin{equation}
\vec{\bf J}=\vec{x}\,{\bf J}^{x}+\vec{y}\,{\bf J}^{y} \quad,\quad \vec{\bf I}=\vec{x}\,{\bf I}^{x}+\vec{y}\,{\bf I}^{y} \ .
\end{equation}
The unit-vectors $\vec{x},\vec{y}$ are related to the orientation of spin (by coupling to $S^x, S^y$), as opposed to the unit-vectors $\hat{\bf x},\hat{\bf y}$ that are related to the spatial orientation of current flow. Formally, $\vec{x},\vec{y}$ and $\hat{\bf x},\hat{\bf y}$ live in different vector spaces, while $\vec{\bf J},\vec{\bf I}$ live in both respective vector spaces at the same time. It is also useful to define:
\begin{equation}
\vec{\theta}=\vec{x}\cos\theta+\vec{y}\sin\theta \quad,\quad
  \hat{\boldsymbol{\theta}}=\hat{{\bf x}}\cos\theta+\hat{{\bf y}}\sin\theta \ .
\end{equation}

We proceed with derivations of detailed and coarse-grained conservation laws for the major types of condensates that we have encountered. We will present only the results for the triplet condensates where approximations are necessary. The exact conservation laws for the two-component condensates are not too tedious to derive analytically, but they are complicated to present and offer no new insight.

\subsection{$S=1$ type-I states}

The type-I states are obtained by naively setting $\boldsymbol{\nabla}\gamma=0$ in various expressions for currents, and relying on $\boldsymbol{\nabla}\alpha \perp \hat{\boldsymbol{\theta}}$ to produce helical spin currents. The gauge-invariant charge current
\begin{equation}\label{ChCurProblem}
{\bf i} = \frac{2\zeta^{2}}{m}\boldsymbol{\nabla}\gamma = 0
\end{equation}
vanishes as expected. Note that the analogous two-component type-I condensate breaks the TR symmetry and has a non-vanishing charge current \cite{Ozawa2012a}, which however still satisfies the local conservation law. The gauge-covariant spin currents obtained from (\ref{CurrentsT1}) are:
\begin{eqnarray}
\vec{\bf J} \!\!&=&\!\! \frac{2\zeta^{2}}{m}\left\lbrack \vec{\theta}\,\boldsymbol{\nabla}\alpha
  -\frac{\sin(2\alpha)}{2}(\vec{z}\times\vec{\theta})\boldsymbol{\nabla}\theta\right\rbrack \\
\vec{\bf I} \!\!&=&\!\! \vec{\bf J}
  +2v\zeta^{2}\biggl\lbrack\cos^{2}\!\alpha\,\vec{\theta}(\hat{{\bf z}}\times\hat{\boldsymbol{\theta}})
  +\sin^{2}\!\alpha\,(\vec{x}\,\hat{{\bf y}}-\vec{y}\,\hat{{\bf x}})\biggr\rbrack \nonumber \ .
\end{eqnarray}
The conservation laws (\ref{CurCons2}) can now be resolved in terms of $\theta$, $\alpha$ and $\zeta$. After some straight-forward algebraic manipulations, one finds:
\begin{eqnarray}\label{CurCons3}
\boldsymbol{\nabla}{\bf J}^{z} \!\!&=&\!\! 4v\zeta^{2}\left(\cos^2\!\alpha\,\boldsymbol{\nabla}\alpha
  +\frac{\sin(2\alpha)}{2}\,\frac{\boldsymbol{\nabla}\zeta}{\zeta}\right)\hat{\boldsymbol{\theta}} \nonumber \\
\vec{\theta}\,\boldsymbol{\nabla}\vec{\bf J} \!\!&=&\!\!
    4v\zeta^{2}\left(\cos^{2}\!\alpha\,\hat{\boldsymbol{\theta}}\boldsymbol{\nabla}\theta
   -\frac{\boldsymbol{\nabla}\zeta}{\zeta}(\hat{\bf z}\times\hat{\boldsymbol{\theta}})\right) \nonumber \\
&& +mv^{2}\zeta^{2}\sin(2\alpha) \\[0.05in]
(\vec{z}\times\vec{\theta})\boldsymbol{\nabla}\vec{\bf J} \!\!&=&\!\! 4v\zeta^{2}\left(
    \frac{\sin(2\alpha)}{2}\boldsymbol{\nabla}\alpha
   +\frac{\boldsymbol{\nabla}\zeta}{\zeta}\sin^{2}\!\alpha\right)\hat{\boldsymbol{\theta}} \nonumber \ .
\end{eqnarray}

The coarse-grained currents (\ref{CurrentsT1}) in the presence of rapid $\alpha$ oscillations:
\begin{equation}\label{CurrentsT1b}
{\bf J}^{z} = -\frac{\zeta^{2}}{m}\,\boldsymbol{\nabla}\theta \quad,\quad
  \vec{\bf J} = \frac{2\zeta^{2}}{m}\,\vec{\theta}\,\boldsymbol{\nabla}\alpha
\end{equation}
obey much simpler coarse-grained conservation laws (\ref{CurCons3}):
\begin{eqnarray}\label{CurCons4}
\boldsymbol{\nabla}{\bf J}^{z} \!\!&=&\!\! 2v\zeta^{2}\,\hat{\boldsymbol{\theta}}\boldsymbol{\nabla}\alpha \\
\vec{\theta}\,\boldsymbol{\nabla}\vec{\bf J} \!\!&=&\!\! 4v\zeta^{2}\left(
   \frac{1}{2}\hat{\boldsymbol{\theta}}\boldsymbol{\nabla}\theta
  -\frac{\boldsymbol{\nabla}\zeta}{\zeta}(\hat{{\bf z}}\times\hat{\boldsymbol{\theta}})\right) \nonumber \\
(\vec{z}\times\vec{\theta})\boldsymbol{\nabla}\vec{\bf J} \!\!&=&\!\! 2v\zeta\,\hat{\boldsymbol{\theta}}\boldsymbol{\nabla}\zeta
  \nonumber \ .
\end{eqnarray}
We can immediately see that $\boldsymbol{\nabla}{\bf J}^z \to 0$ on fairly short length-scales if we keep $\hat{\boldsymbol{\theta}}$ and $\hat{{\bf z}}\times\boldsymbol{\nabla}\alpha$ parallel to each-other to minimize the Rashba energy. Therefore, ${\bf J}^z$ should have no sources or drains. Only current loops in the form of quantized vortices can make ${\bf J}^z$ finite.

Substituting the divergences of (\ref{CurrentsT1b}) into (\ref{CurCons4}) yields:
\begin{eqnarray}\label{CurCons5}
-\frac{2\zeta}{m}(\boldsymbol{\nabla}\zeta)(\boldsymbol{\nabla}\theta)-\frac{\zeta^{2}}{m}\nabla^{2}\theta \!\!&=&\!\!
  2v\zeta^{2}\,\hat{\boldsymbol{\theta}}\boldsymbol{\nabla}\alpha\to 0 \nonumber \\
\frac{4\zeta}{m}(\boldsymbol{\nabla}\zeta)(\boldsymbol{\nabla}\alpha)+\frac{2\zeta^{2}}{m}\nabla^{2}\alpha \!\!&=&\!\!
  2v\zeta^{2}\left( \hat{\boldsymbol{\theta}}\boldsymbol{\nabla}\theta
  -\frac{2}{\zeta}(\hat{{\bf z}}\times\hat{\boldsymbol{\theta}})\boldsymbol{\nabla}\zeta\right) \nonumber \\
\frac{2\zeta^{2}}{m}(\boldsymbol{\nabla}\theta)(\boldsymbol{\nabla}\alpha) \!\!&=&\!\!
  2v\zeta\,\hat{\boldsymbol{\theta}}\boldsymbol{\nabla}\zeta
\end{eqnarray}
in the ground state ($\hat{\boldsymbol{\theta}}\boldsymbol{\nabla}\alpha \to 0$). These consequences of spin-current conservation look particularly simple in the regions far away from any vortex singularities, where the spatial variations of the order parameter magnitude $\zeta$ can be neglected:
\begin{eqnarray}\label{CurCons6}
\nabla^{2}\theta \!\!&=&\!\! -2mv\hat{\boldsymbol{\theta}}\boldsymbol{\nabla}\alpha \to 0 \\
\nabla^{2}\alpha \!\!&=&\!\! mv\,\hat{\boldsymbol{\theta}}\boldsymbol{\nabla}\theta \nonumber \\
(\boldsymbol{\nabla}\theta)(\boldsymbol{\nabla}\alpha) \!\!&=&\!\! 0 \ . \nonumber
\end{eqnarray}

It is instructive to examine the vortex configurations that we discovered earlier through the lenses of the above equations. For example, the $Q=+1$ chiral vortices of gradually winding $\theta$ can be arranged to have co-rotating vectors $\boldsymbol{\nabla}\theta$ and $\hat{\boldsymbol{\theta}}$ on the closed loops around the singularity. The second equation of (\ref{CurCons6}) predicts that helical currents $\boldsymbol{\nabla}\alpha$ must have distributed radially-symmetric sources in that environment. This equation can be readily solved by the Gauss' theorem, and yields precisely the source-like configuration $\boldsymbol{\nabla}\alpha = mv \hat{\bf r} = -mv \hat{\bf z} \times \hat{\boldsymbol{\theta}}$ that is made optimal by the Rashba spin-orbit coupling. We anticipated such current patterns near the $Q=+1$ vortices of the metastable clusters. The other two equations of (\ref{CurCons6}) are also consistent with this current pattern. If we attempt to solve these equations in the vicinity of an ordinary $Q=-1$ chiral antivortex with \emph{gradually} varying angle $\theta$, the outcome of the second equation is the $\boldsymbol{\nabla}\alpha$ pattern given by (\ref{Da}), which is also optimal with respect to the Rashba spin-orbit coupling. We noted that this pattern cannot correspond to a gradient of a scalar ($\alpha$), but the equations of motion are not very sensitive to this issue given that they are classical in spirit. It is only the last equation of (\ref{CurCons6}) that catches a problem with this configuration: it cannot be satisfied at any finite distance from the vortex core, where $\boldsymbol{\nabla}\theta$ is finite. In fact, this hints the resolution of the problem that we already found: $\boldsymbol{\nabla}\theta$ must be made zero in the bulk by compressing all of its winding into narrow strings attached to vortices. The above coarse-grained equations break down at the length-scales of the string thickness.

Therefore, we can see all fundamental features of chiral vortices through the current conservation laws without trying to solve them in detail. Helical vortices are not that easy. The winding of $\alpha$ in the $\theta=\textrm{const}$ environment trivially satisfies the last two equations of (\ref{CurCons6}), but seemingly violates the first equation. It should be kept in mind, however, that these equations are coarse-grained, so we shouldn't take seriously their discrepancy near the singularity, where $\boldsymbol{\nabla}\alpha$ has a rapid circulation. On the other hand, $\boldsymbol{\nabla}\alpha$ has a slowly varying circulating component inversely proportional to the distance from the core, so it changes direction with respect to $\hat{\boldsymbol{\theta}}$ in different regions far away from the core and necessitates a distribution for sources and drains of chiral currents $\boldsymbol{\nabla}\theta$. These currents cannot remain strictly zero, and the problem becomes very complicated. We cannot solve the conservation equations, but may still rest assured that helical vortices cost a logarithmically divergent energy with system size. We have discovered \emph{one} configuration of a helical vortex that costs a logarithmic energy. This might not satisfy the conservation laws as an equilibrium state. Nevertheless, the true equilibrium state with given boundary condition at the vortex core is the minimum-energy state, so it cannot cost more energy than the one state we found.

\subsection{$S=1$ type-II states}

The gauge-covariant charge $\bf i$ and spin $\bf I$ currents in type-II states (\ref{OP3T2}) are:
\begin{eqnarray}
{\bf i} &=& \frac{2\zeta^{2}}{m}\boldsymbol{\nabla}\gamma+2v\zeta^{2}\sin(2\alpha)(\hat{{\bf z}}\times\hat{\boldsymbol{\theta}}) \\
{\bf I}^{x} &=& \frac{2\zeta^{2}}{m}\sin(2\alpha)\cos\theta\,(\boldsymbol{\nabla}\gamma)
    +2v\zeta^{2}\biggl\lbrack\cos^{2}\alpha\,\sin\theta\,\hat{\boldsymbol{\theta}}-\hat{{\bf y}}\biggr\rbrack \nonumber \\
{\bf I}^{y} &=& \frac{2\zeta^{2}}{m}\sin(2\alpha)\sin\theta\,(\boldsymbol{\nabla}\gamma)
    -2v\zeta^{2}\biggl\lbrack\cos^{2}\alpha\cos\theta\,\hat{\boldsymbol{\theta}}-\hat{{\bf x}}\biggr\rbrack \nonumber \\
{\bf I}^{z} &=& -\frac{2\zeta^{2}}{m}\cos^{2}\alpha\,\boldsymbol{\nabla}\theta \nonumber
\end{eqnarray}
where we used (\ref{ThetaHat1}). Substituting this into (\ref{CurCons2}) gives us:
\begin{eqnarray}
\frac{1}{mv}\boldsymbol{\nabla}{\bf I}^{x} &=& \frac{2\zeta^{2}}{m}\cos^{2}\alpha\,\partial_{x}\theta \\
\frac{1}{mv}\boldsymbol{\nabla}{\bf I}^{y} &=& \frac{2\zeta^{2}}{m}\cos^{2}\alpha\,\partial_{y}\theta \nonumber \\
\frac{1}{mv}\boldsymbol{\nabla}{\bf I}^{z} &=& \frac{2\zeta^{2}}{m}\sin(2\alpha)\,\hat{\boldsymbol{\theta}}\boldsymbol{\nabla}\gamma \nonumber \ .
\end{eqnarray}
After some algebra, we obtain the conservation laws for type-II condensates in terms of $\zeta, \gamma, \theta$:
\begin{eqnarray}\label{CurCons7}
&&  ~~~~~~~~~~~~~~~~~ \zeta^{2}\,\nabla^{2}\gamma + 2\zeta(\boldsymbol{\nabla}\zeta)(\boldsymbol{\nabla}\gamma) = 0 \vspace{-0.05in} \\
&& \sin(2\alpha)\left(\hat{\boldsymbol{\theta}}\boldsymbol{\nabla}\theta
     -2\frac{\boldsymbol{\nabla}\zeta}{\zeta}\hat{{\bf z}}\times\hat{\boldsymbol{\theta}}\right)
   = 2\cos(2\alpha)(\hat{{\bf z}}\times\hat{\boldsymbol{\theta}})\boldsymbol{\nabla}\alpha \nonumber \\
&&  \sin\alpha\,(\boldsymbol{\nabla}\theta)(\boldsymbol{\nabla}\gamma) = mv\,\cos\alpha\,\left(-(\hat{{\bf z}}\times\boldsymbol{\nabla}\theta)
  + \frac{\boldsymbol{\nabla}\zeta}{\zeta}\right)\hat{\boldsymbol{\theta}} \nonumber \\
&& \!\!\!\!\!\!\!\! -2\zeta\cos^{2}\alpha(\boldsymbol{\nabla}\theta)(\boldsymbol{\nabla}\zeta)-\zeta^{2}\cos^{2}\alpha\nabla^{2}\theta = 
  mv\zeta^{2}\sin(2\alpha)\,\hat{\boldsymbol{\theta}}\boldsymbol{\nabla}\gamma \nonumber
\end{eqnarray}
The first two equations follow from the conservation of charge currents (\ref{ChargeCons}) and $z$-projection spin currents ${\bf J}^z$ in equilibrium.

In the optimum type-II state with $\alpha=\frac{\pi}{4}$, far away from vortex cores $\boldsymbol{\nabla}\zeta \to 0$, the above equations simplify further:
\begin{eqnarray}
\nabla^{2}\gamma\!\!&=&\!\!0 \\
\hat{\boldsymbol{\theta}}\boldsymbol{\nabla}\theta\!\!&=&\!\!0 \nonumber \\
(\boldsymbol{\nabla}\theta)(\boldsymbol{\nabla}\gamma)\!\!&=&\!\!
  -mv\,(\hat{{\bf z}}\times\boldsymbol{\nabla}\theta)\hat{\boldsymbol{\theta}} \nonumber \\
\nabla^{2}\theta\!\!&=&\!\!-2mv\,\hat{\boldsymbol{\theta}}\boldsymbol{\nabla}\gamma\to0 \ . \nonumber
\end{eqnarray}
The first equation prohibits sources and drains for charge currents as expected. The last equation prohibits sources and drains of chiral spin currents $\boldsymbol{\nabla}\theta$ in the states with minimized Rashba energy ($\boldsymbol{\nabla}\gamma \perp \hat{\boldsymbol{\theta}}$ in the $S=1$ systems). The second and third equations are satisfied in the uniform type-II state, as well as in the vicinity of an isolated conventional $Q=+1$ chiral vortex. However, they object the ordinary $Q=-1$ chiral antivortices. The general winding of $\theta$ is consistent with these equations only if it is compressed into strings.

\subsection{Short-scale density modulations}\label{secDensityMod}

Here we reveal and discuss the existence of spin density modulations at length-scales below $(mv)^{-1}$ in some states that we have discussed. These modulations are necessary in various circumstances in order to satisfy the detailed conservation laws. However, the coarse-grained conservation laws at length-scales beyond $(mv)^{-1}$ are consistent with all naive uniform condensates and large-scale vortex structures.

The simplest example of this phenomenon is the uniform $S=1$ type-I order parameter (\ref{OP3T1}). If the time derivatives are left out of (\ref{CC}), we find for this order parameter:
\begin{equation}\label{CCuniform}
\partial_i I_i -i\lbrack\mathcal{A}_i,I_i\rbrack = - m v^2 \zeta^2 \sin(2mvx) S^y \ ,
\end{equation}
so it fails to satisfy the detailed equilibrium spin-current conservation laws. The non-zero value on the right-hand side is the rate at which the spin density wants to relax to a different configuration. We see that relaxation occurs only at length-scales of the order of $(mv)^{-1}$. The final static state exhibits spatial modulations with a period $\sim (mv)^{-1}$. There is no net relaxation on large length-scales, since the right-hand side averages to zero. We could naively expect spontaneous TR-symmetry breaking by the ensuing spin density-wave, but we cannot prove it without solving the equations in detail. We can only rest assured that there is no ferromagnetic spin polarization at large length-scales. It is also important to note that there is a special combination of the couplings ($t_{\textrm{t}}$, $a$) in (\ref{LG3}) or (\ref{NonMin}) which resets the right-hand side of (\ref{CCuniform}) to zero. These special values of the coupling constants cancel out exactly the ``diamagnetic'' term $\mathcal{A}^2/2m$ in the action (\ref{LG3}), leaving behind the pure Rashba spin-orbit coupling $v \hat{\bf z} ({\bf S} \times {\bf p})$. The ensuing static state requires no relaxation in terms of spin currents and spontaneous TR-symmetry breaking.

Similar short-scale modulations occur in the vicinity of vortices, even when the ground state is free of modulations. The order parameters of static and smooth large-scale vortex structures generally violate the detailed current conservation laws. This can be seen by the same analysis that led to (\ref{CCuniform}). The energy of modulated vortex structures is only lower than that we estimated from the unmodulated ones. A pursuit of detailed modulation current patterns is beyond the scope of this paper.


%

\end{document}